\documentclass[a4paper,10pt]{elsarticle}
\usepackage[utf8]{inputenc}
\usepackage{amssymb}
\usepackage{amsmath}
\usepackage{amsfonts}
\usepackage{geometry}
\usepackage{graphicx}
\usepackage{tabularx}
\usepackage{longtable}
\usepackage{mathrsfs}
\usepackage[unicode,pdfauthor={S. Shashank et al.}, pdftitle={f-mode oscillations of compact stars}]{hyperref}
\hypersetup{colorlinks,urlcolor=blue,citecolor=blue,linkcolor=blue}

\begin{document}

\begin{frontmatter}

\date{}

\title{$f$-mode oscillations of compact stars with realistic equations of state in dynamical spacetime}

\author[1, 2]{Swarnim Shashank}
\author[1, 3]{Fatemeh Hossein Nouri}
\author[1]{Anshu Gupta}

\address[1]{Inter-University Centre for Astronomy and Astrophysics, Post Bag 4, Ganeshkhind, Pune 411 007, India}
\address[2]{Center for Field Theory and Particle Physics and Department of Physics, Fudan University, 200438 Shanghai, China}
\address[3]{Center for Theoretical Physics, Polish Academy of Sciences, Al. Lotnikow 32/46, 02-668 Warsaw, Poland}

\begin{abstract}
In this study, we perform full three dimensional numerical relativity simulations of non-rotating general relativistic stars. 
Extending the studies for polytropic equation of state, we investigate the accuracy and robustness of numerical scheme on measuring fundamental ($f$)-mode frequency for realistic equations of state (EoS).
We use various EoS with varying range of stiffness and numerically evolve perturbed stellar models for several mass configurations (in the range of $1.2 - 2.0$ $M_{\odot}$) for each of these EoS.
Using the gravitational waveform obtained from the simulations we extract the $f$-modes of the stars. The obtained results are tested against the pre-existing perturbation methods and find good agreement. Validity and deviation of universal relations have been carried out and are compared with earlier results under Cowling approximation as well as perturbative approaches. We also show that even using perturbed single star simulations can provide good agreement with $f$-modes extracted from inspiral phase of binary neutron star simulations which are computationally more expensive.
\end{abstract}


\end{frontmatter}

\section{Introduction}
\label{sec:intro}

Asteroseismology that enables us to understand the interior structure of neutron stars is coming to the forefront, with added tools of gravitational wave astronomy applied in detecting gravitational wave signal from two possible binary neuron star mergers (GW170817 \cite{GW170817} \& GW190425 \cite{GW190425}) and electromagnetic observations like NICER data. This assists in resolving degenerate parameters, puts stricter bounds on the internal composition of the star \cite{miller2019:ApJ...887L..24M, bogdanov2019a:ApJ...887L..25B, bogdanov2019b:ApJ...887L..26B, raaijmakers2020:ApJ...893L..21R, Jiang2020:ApJ...892...55J} and tests universal relations (UR).

Some of the parameters of neutron stars which could be inferred and estimated through multi-messenger astronomy are the mass, radius, spin, tidal deformability based on the mode frequencies extracted from the signals. While the joint mass, radius estimates using NICER data has put limits on the possible equations of state (EoS) which describes the internal composition \cite{miller2019:ApJ...887L..24M, bogdanov2019a:ApJ...887L..25B, bogdanov2019b:ApJ...887L..26B}, incorporating tidal deformability information based on the GW170817 constraints further, the bounds on the permissible EoS \cite{raaijmakers2020:ApJ...893L..21R, miller2020:ApJ...888...12M}. 

Some of the earlier works on gravitational asteroseismology includes Ref.~\cite{andersson_kokkotas1998, gabrielle1998:PhysRevD.58.124012, pons2002:PhysRevD.65.104021, Benhar2004:PhysRevD.70.124015} (see Ref.~\cite{Kokkotas_1996}
for the pulsations of relativistic stars and references therein). 
Through multiple studies it has been shown that the fundamental oscillation mode ($f$-mode) of the star has a strong resonance with the orbital frequencies closer to merger of a coalescing binary neutron star. Resonant tidal excitation of oscillation modes in merging binary neutron stars has been carried out in Ref.~\cite{Lai1994, Lai2006, Xu-Lai2017}. Other oscillation modes of the star, like $p$ and $g$-modes, may also become relevant throughout the inspiral, due to nonlinear coupling to the tide, as discussed in Ref.~\cite{Weinberg_2013, Weinberg_2016, Zhou_2017, Nouri:2021}.

Phenomenological models connecting tidal deformability with frequency have been discussed in Ref.~\cite{Andersson19:arxiv1905, Andersson20:PhysRevD.101.083001}. For GW170817 data, $f$-mode frequency have been estimated in Ref.~\cite{hinderer_nature}  by using an $f$-mode tidal model (fmtidal) in the frequency domain described in Ref.~\cite{Schmidt_Hinderer:2019_PhysRevD}. Contribution due to spin during the binary neutron star merger on $f$-modes and dynamical tides have been recently carried out in Ref.~\cite{PhysRevD.101.123020, steinhoff2021spin}.
Gravitational-wave asteroseismology with $f$-modes from neutron star binaries at the merger phase has been also carried out recently  \cite{Tjonnie-li:arxiv2012}, where they use NR BNS simulations carried out in Ref.~\cite{Rezzolla_2016:PhysRevD.93.124051, Dietrich_2017a:PhysRevD.95.024029, Dietrich_2017b:PhysRevD.95.044045} and find less than one percent difference between BNS merger frequency (based on the merger peak amplitude) and $f$-modes computed for isolated neutron stars. 
Frequency deviations in the universal relations of isolated neutron stars and postmerger remnants have been discussed in Ref.~\cite{Nick-etal:arxiv2102}, whereas UR for damping time have been carried out in Ref.~\cite{damping_nick_2018}.   
Universal relations among compactness ($\mathcal{C}=M/R$), moment of inertia ($I$) and tidal deformability, and their validity and deviations have been studied in Ref.~\cite{Maselli2013:PhysRevD.88.023007, ChanUR:PhysRevD.90.124023, Chirenti:2015PRD, Yagi:2016qmr} for the various choices of equations of state (see Ref.~\cite{Yagi:2016bkt} for review).

Series of studies have been performed in computing oscillation modes of neutron star using linear perturbation theory \cite{Lindblom:1983ps, Detweiler_1985, Chirenti:2015PRD}.
These studies are based on the standard approach laid out in Thorne and Campolattaro (1967) \cite{1967ApJ...149..591T} to solve the stellar perturbation equations in curved spacetime by applying the appropriate boundary conditions. The $f$-mode oscillations have also been studied using numerical simulations, employing the cowling approximation i.e. by evolving hydrodynamic equations in the fixed background of general relativistic spacetime for a single perturbed neutron star \cite{Font_2000, font_rotating, Shibata2004, Kastaun_2010_CA:PhysRevD.82.104036, Chirenti:2015PRD, Hebert-2018}, also in full GR to study bar mode instability~\cite{DePietri2014}, as well as under conformally flatness condition (CFC) \cite{Dimmelmeier2006:MNRAS, Bucciantini_2011:aap, Pili_2014:mnras}, where the 3-metric is assumed to be conformally flat and the spacetime dynamics is coupled with fluid dynamics.

Recently, Rosofsky et al. \cite{rosofsky} carried out studies in the fully dynamical spacetime and extracted the fundamental modes for the polytropic equations of state. There have been studies  
using piece-wise poly-tropic EoS to impose constraints on the neutron star structure \cite{Bauswein_2020_PhysRevLett.125.141103, miller2020:ApJ...888...12M} or using the parametrised EoS with continuous sound speed \cite{Boyle_nick_2020_PhysRevD.102.083027} as well.

For Asteroseismology studies in curved spacetime, it is important to utilize the full general relativistic hydrodynamical simulations to study neutron stars' EoS. The current standard nuclear EoS used in numerical simulations (tables in https://stellarcollapse.org for instance) are in the tabulated form. This means that the status of the fluid is identified by three quantities $(\rho,T,Y_e)$, where $\rho$ is the baryonic density, $T$ is the temperature and $Y_e$ is the electron fraction. Obviously, the accuracy of the numerical simulations with this type of EoS is affected by the resolution of the three dimensional $(\rho,T,Y_e)$ table. Moreover, the non-smoothness features of various physical quantities in such EoS can be another source of numerical errors for long-term evolutions. Therefore, it is crucial to perform numerical simulations with tabulated EoS to examine the accuracy of the current general relativistic hydrodynamical codes for measuring the oscillation mode frequencies, and test their results against the analytical results. 
In the current work, we present a few numerical tests to investigate the accuracy challenges mentioned above, and then we extract the $f$-mode frequencies by evolving nonrotating neutron star in the dynamical spacetime while considering different tabulated nuclear EoS. 

We also probe the possibility of using these single star simulations to study the tidal effects expected during a binary inspiral. This is done by perturbing the initial density field of the star with the dominant quadrupolar term in the spherical harmonics expansion of the tidal field $Y_{22}$. Single star simulations are computationally less expensive to carry out, providing the possibility to reach higher resolutions in the future which may be important for studying higher order harmonics. We, further, study the validity of our results using the URs associated with tidal deformability and stellar parameters.

This is a step towards extending our study incorporating spin effects which we shall report in the follow up work. Our work provides leeway for future studies of tidal effects of rotating stars in an inspiral and also post-merger remnants with differential rotation.

In Sec.~\ref{sec:basic equations} we briefly outline mathematical and numerical framework for general relativistic hydro-dynamical system, the initial setup and matter configuration as described by a set of equations of state under consideration. 
In Sec.~\ref{sec:num_tests} we discuss the accuracy challenges of mode frequency measurements for tabulated EoS, and we present the results of the several numerical tests to demonstrate the accuracy of our measurement. 
We analyse our simulation data and describe the result findings in Sec.~\ref{sec:results}. We compute fundamental mode ($f$-mode frequency), its relations as a function of compactness and mass, then compare our fits with some of the recently carried out works. Our findings match well with the universal relations described with a deviation less than of a few percent. Sec.~\ref{sec:conclusions} discusses our results and conclusions.


\section{Basic Numerical Framework} 
\label{sec:basic equations}
We solve the Einstein equations using Numerical Relativity methods of 3+1 decomposition of the spacetime.
\begin{equation}
\label{eqn:efe}
    G_{\mu\nu} \equiv R_{\mu\nu} - \frac{1}{2}g_{\mu\nu}R = 8 \pi T_{\mu\nu}
\end{equation}
\begin{equation}
\label{eqn:line_element}
    ds^2 = ( - \alpha^2 + \beta_i \beta^i ) dt^2 + 2 \beta_i dx^i dt + \gamma_{ij} dx^i dx^j
\end{equation}
\cite{Alcubierre_2008, Rezzolla_Zanotti_2013, Baumgarte_Shapiro_2010, Shibata_2015}
here, $\alpha$ is the lapse function, $\beta ^i$ is the shift vector and $\gamma_{ij}$ is the spatial metric. The units used are $G = c = 1$ unless mentioned otherwise.

\subsection{Dynamical evolution}
\label{subsec:evolve}
The sapcetime evolution is achieved using the Baumgarte-Shapiro-Shibata-Nakamura-Oohara-Kojima (BSSNOK) formalism \cite{oohara_kojima, shibata_nakamura, baumgarte_shapiro, alcubierre_bruegmann} which is a conformal formulation of the ADM equations \cite{adm1, adm2}. The (1+log) and Gamma-driver gauge conditions are adopted for the evolution of lapse and shift \cite{Alcubierre_2008, Baiotti_Rezzolla_2017_review}. We use the \textsc{MacLachlan} code \cite{ml_bssn} for evolution of spacetime variables, which is a publicly available code in the \textsc{EinsteinToolkit} \cite{etk2019, cactus, carpet1, carpet2} suite. A Kreiss-Oliger dissipation \cite{Rezzolla_Zanotti_2013, Alcubierre_2008, Shibata_2015} is added to spacetime variables for removing high frequency noise.\\
To model Neutron stars, a relativistic perfect fluid is assumed. The conservation equations for the energy-momentum tensor $T_{\mu\nu}$ and the matter current density $J_{\mu}$, are solved numerically after being recast into a flux-conservative formulation \cite{Rezzolla_Zanotti_2013, Font_living_review, Baiotti_Rezzolla_2017_review}
\begin{equation}
\label{eqn:hydro}
    \nabla_{\mu} J^{\mu} = 0, \nabla_{\mu} T^{\mu\nu} = 0.
\end{equation}

The system of equations is complete with an Equation of State of the type $p = p(\rho, Y_e, T)$, which for our models is described in the Sec.~\ref{subsec:EoS}. (For more details readers can refer to Ref.~\cite{Rezzolla_Zanotti_2013, Font_living_review, Baiotti_Rezzolla_2017_review, Shibata_2015}.) We carry out the hydrodynamics using the publicly available \texttt{WhiskyTHC} code \cite{thc1, thc2, thc3} which works within the \textsc{EinsteinToolkit} framework and uses high-resolution shock capturing methods. For the time integration, method of lines is used with fourth-order Runge-Kutta methods. The fifth-order MP5 flux-reconstruction method is used along with Harten-Lax-van Leer-Einfeldt (HLLE) Riemann solver.

\subsection{Initial data}
\label{subsec:id}

We use a perturbed Tolman–Oppenheimer–Volkoff (TOV) star for the initial data.
The initial data is generated using the \texttt{PizzaTOV} thorn for both polytropic and tabulated equations of state. A perturbation is added for the density as:
\begin{equation}\label{eqn:perturbation}
    \delta \rho = \mathcal{A} \rho (r / R) Y_{2 2},
\end{equation}
where $\delta \rho$ is the perturbed density, $\rho$ is the density, $\mathcal{A}$ is the perturbation amplitude, which we set to $0.01$ to introduce a small perturbation, $r$ is the radial distance from centre of the star and $R$ is the radius of the star. We use the $(2, 2)$ eigenfunction which is expected to be similar to tidal interactions of binary neutron stars \cite{rosofsky, hinderer_nature}. For all the models we choose the artificial background atmosphere density as $\rho_{atm} = 10^{-14}~M^{-2} \approx 6.17 \times 10^{3} g~cm^{-3}$.\\
The simulations are performed at minimum grid spacing of $0.105M \approx 155m$. For the tabulated EoS $1.4M_{\odot}$ cases we also perform simulations at $0.07M$ and for polytrope at $0.06M$. The convergence test for the DD2 equation of state for three different resolutions is given in Sec.~\ref{sec:dd2_conv}.

\subsection{Equations of State}
\label{subsec:EoS}

\begin{figure}
    \centering
    \includegraphics[width=0.55\textwidth]{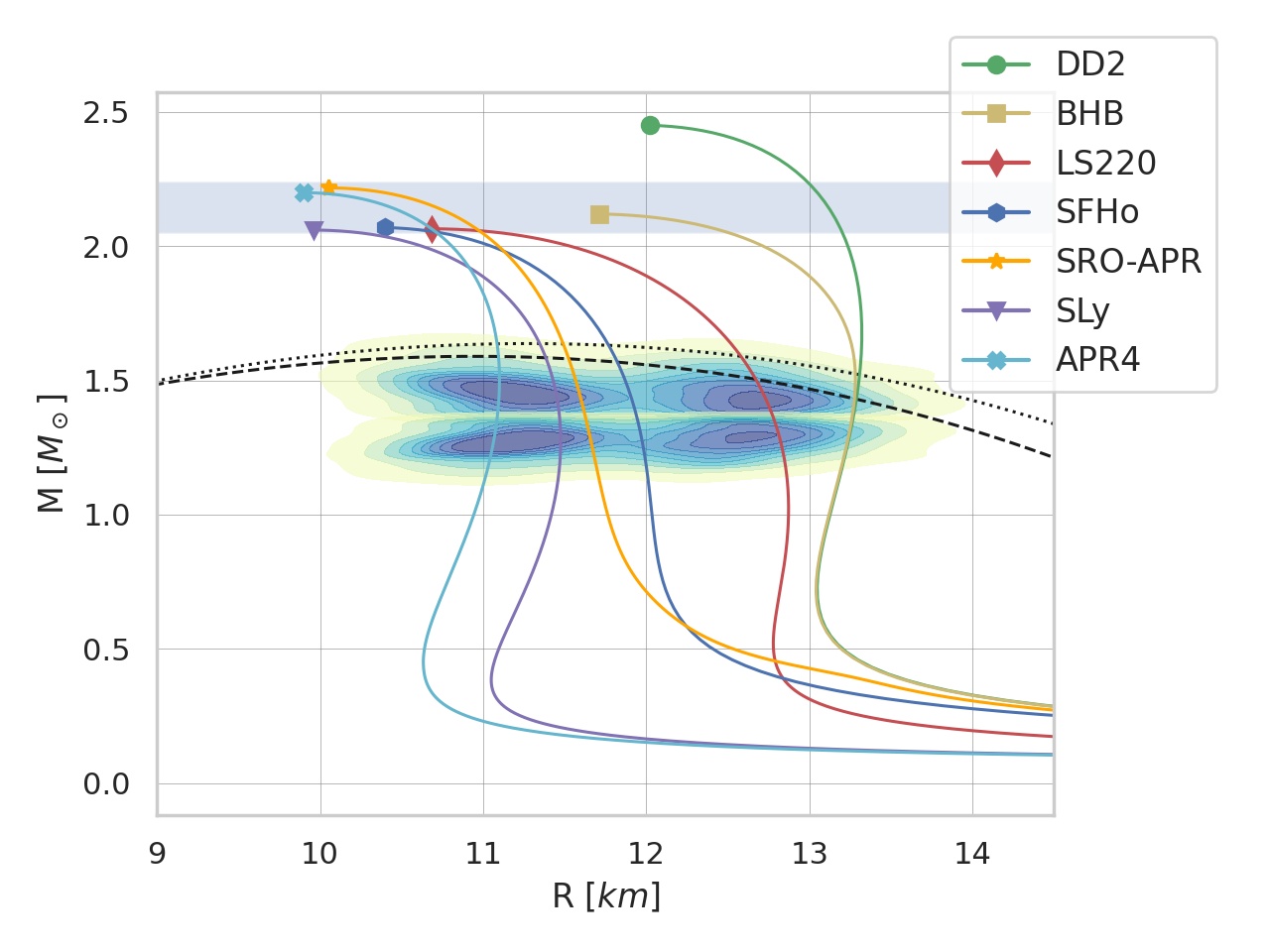}
    \caption{A comparison of all the equation of states used. The dotted line represents our polytropic model with $\kappa = 100$ and the dashed line represents the polytropic model $\kappa=94.29$ (see Eq.~\ref{eqn:polytropic}). The shaded region is the range of observed mass of pulsar $J0740+6620$ \cite{Cromartie2020}. Purple-Blue region in the middle is the parametrised EoS $M-R$ relation obtained from $GW170817$ \cite{EoS_GW170817}.}
    \label{fig:EoS_compare}
\end{figure}

We use several finite-temperature, composition dependent nuclear-theory based equations of state (Fig.~\ref{fig:EoS_compare}). Three of them are based on relativistic mean field (RMF) models.
These equations of state are publicly available in tabulated form at https://stellarcollapse.org. The resolution of tables used are around 250-300 points.

\begin{enumerate}

\item LS220~\cite{Lattimer:1991} (Lattimer \& Swesty EoS with the incompressibility K = 220 MeV): contains neutrons, protons, alpha particles and heavy nuclei.
It is based on the single nucleus approximation for heavy nuclei. LS220 has been widely used in many supernova simulations.

\item DD2~\cite{Hempel_2012}: contains neutrons, protons, light nuclei such as deuterons, helions, tritons and alpha particles and heavy nuclei. 
DD2 is an RMF with a density-dependent nucleon-meson
coupling for treating high density nuclear matter. 

\item SFHo~\cite{Steiner_2013}: Another RMF, and similar to DD2 it contains neutrons, protons, light nuclei
such as deuterons, helions, tritons and alpha particles
and heavy nuclei.
However, the RMF parameters are tuned to fit the NS mass-radius observation.

\item BHB~\cite{Banik_2014}: Another RMF similar to DD2 and SFHo with the same particle composition, but BHB EoS additionally includes $\Lambda$ hyperons and hyperon-hyperon interactions allowed by $\phi$ mesons.

\item SLy~\cite{Chabanat:1997un}: contains only protons, neutrons and electrons. It is developed out of a refined Skyrme-like effective potential, originating from the shell-model description of the nuclei.

\item APR4: is the complete version of a four-realistic-EoS series developed by Akmal, Pandharipande and Ravenhall~\cite{akmal1998}, named APR1 through 4, obtained from potentials resulting from fits to nucleon-nucleon scattering.
APR4 includes the relativistic corrections and the
three nucleon interaction potential additionally, which makes it more complete in comparison with the other APRs. This EoS is commonly used in neutron star simulations,
as it occurs to be compatible with astronomical
observations.

\item SRO-APR~\cite{Schneider2019}: is based on APR potential model. However, similar to Lattimer \& Swesty EoS, it assumes compressible liquid droplet model of nuclei. 
It contains neutrons, protons, a single type of heavy nucleus plus alpha particles representing light nuclei.

\end{enumerate}

For testing our methods, we also use a polytropic equation of state model
\begin{equation}
\label{eqn:polytropic}
    p = \kappa \rho^{\Gamma}.
\end{equation}
We use $\Gamma = 2$ and choosing accordingly the values of $\kappa$ and central density, we create models for stars of masses $1.35M_{\odot}$ and $1.4M_{\odot}$. The black dashed line in Fig.~\ref{fig:EoS_compare} represents the M-R relation of a polytrope with $\kappa = 94.29$. This parameter is applied to create the model of the star with $1.35M_{\odot}$. The brown dotted line in Fig.~\ref{fig:EoS_compare} represents M-R relation for a polytropic star with $\kappa = 100$ for a star with the mass of $1.4M_{\odot}$. These configurations are considered apart from the other realistic EoSs to compare with the literature and the earlier studies which considered polytropes.
We evolve the polytropic models as well as the \texttt{SLy} and \texttt{APR4} models using a hybrid equation of state approach \cite{takami_hybrid, figura2020hybrid} viz. a gamma-law correction for finite temperature with $\Gamma_{th} = 2$. 
The evolution equation of state becomes:
\begin{equation}
\label{eqn:gamma_law}
    P ( \rho, \epsilon ) = P_{\mathrm{cold }} ( \rho ) + \rho [ \epsilon - \epsilon_{\mathrm{cold }} ( \rho ) ] ( \Gamma_{\mathrm{th }} - 1 )
\end{equation}
We find that the results obtained from the hybrid equation of state method for our $1.4M_{\odot}$ polytrope to be consistent with the results from Ref.~\cite{rosofsky}.

\subsection{Extracting the $f$-mode frequencies}
In order to compute the $f$-modes, we use the Fourier transform of the time series data for the gravitational waveform generated during the evolution of each simulation (except for Cowling case where the frequency of $f$-mode is extracted from Fourier transform of central density).  For our analysis we consider the waveform that is extracted at the radius $r = 10M$ for the $l=2$, $m=2$ mode of the $\Psi_4$ data, which is calculated using the Newman-Penrose formalism \cite{newman_penrose_gw,bishop_rezzolla_gw_extraction}. We choose this radius since it has the least amount of noise among all the considered extraction radii while the computed fundamental mode stays the same as could be seen in Fig.~\ref{fig:convergence_test} (bottom right).


\section{Numerical accuracy tests}
\label{sec:num_tests}

Measuring the frequency of the oscillation modes from a fully general relativistic hydrodynamic evolution can encounter several numerical challenges. 
For an accurate frequency measurement, the space-time and hydro evolutions are supposed to be done accurately with a reasonably-high grid resolution within the convergence regime. 
Particularly, for tabulated equations of state, the accuracy might be highly affected by the resolution of the hydro variables from the original EOS table (the number of the divisions over the density range for instance in our case). This accuracy limit appears specifically in the first derivatives' computation during each time step of the hydro evolution~\cite{Kastaun-2021}.
The reconstruction algorithm can be another possible source of errors for these measurements; Despite the fact that all commonly-used reconstruction methods such as WENO, PPM, etc. formally converge at second order in a standard finite volume/difference scheme, though the choice of the numerical method to reconstruct data at cell faces is found to be critical to correctly capture the stellar oscillations, and some of these methods often have significantly smaller errors compared to other ones~\cite{rosofsky, thc2}.

To address all these accuracy concerns, we designed several numerical tests including convergence and Cowling tests to investigate the accuracy of our simulations' setup and numerical methods used for frequency measurements. The results of these tests prove that our numerical setup and methods provide a good accuracy and robustness for $f$-mode frequency studies.

\subsection{Testing $f$-mode frequency in Cowling approximation}

\begin{figure}
    \centering
    \includegraphics[width=0.55\textwidth]{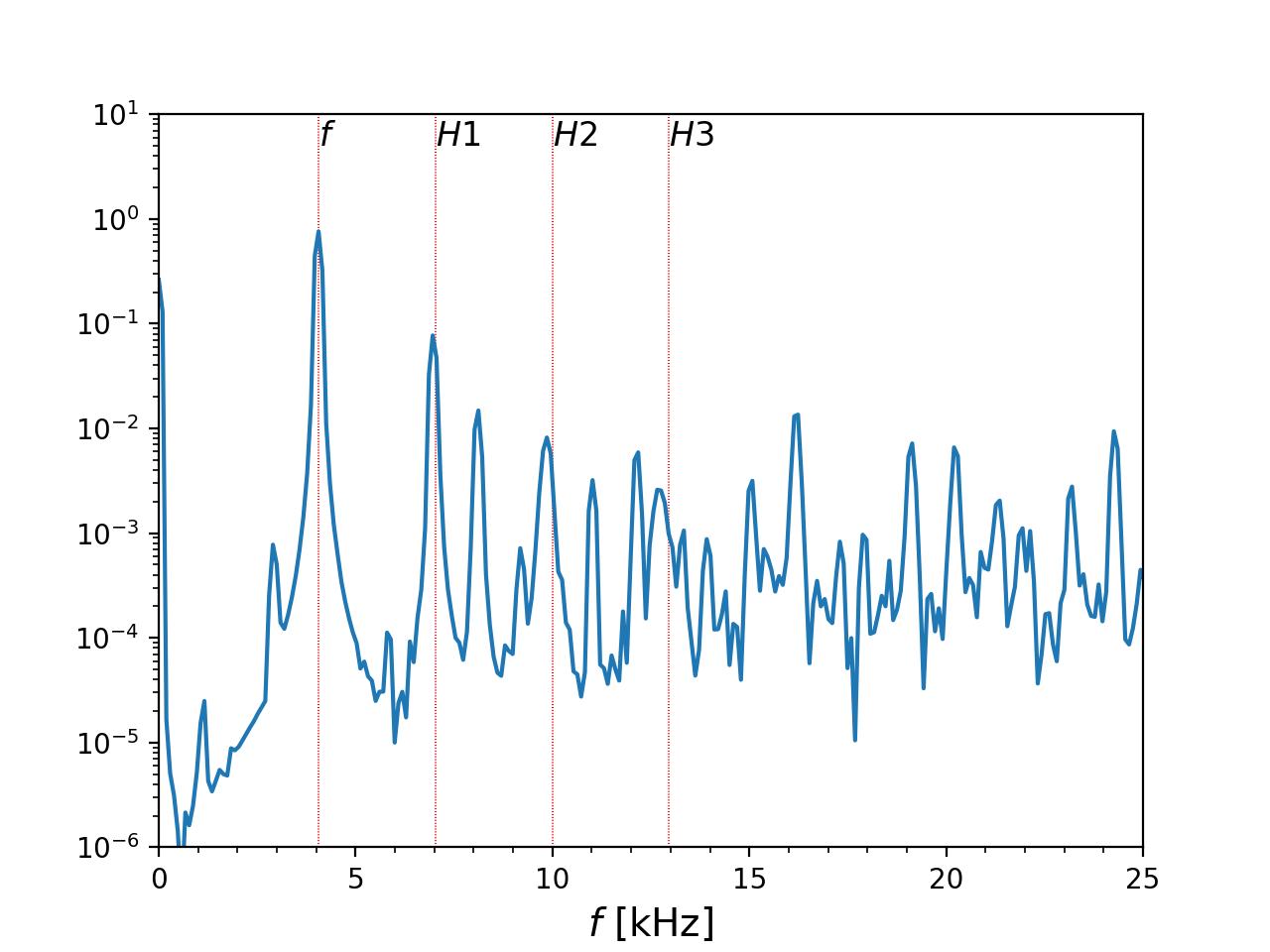}
    \caption{Test simulation in Cowling approximation to check against analytical values (red lines) of $f$-mode frequencies. We find that the simulation retrieves accurate frequencies for $f$-mode and the first harmonic.}
    \label{fig:ls220_cowling}
\end{figure}

As a part of the accuracy investigations, we test the hydrodynamic evolution of our numerical setup verses the analytical results from the linear perturbation theory. We run a test simulation in the Cowling approximation i.e. with spacetime evolution turned-off. In the perturbation theory, the Cowling approximation is taken into account by neglecting the perturbation of the metric components. The mode frequencies are derived by solving a set of coupled ODEs for hydro perturbation equations by applying appropriate boundary conditions (see Ref.~\cite{Finn_1988,Nouri:2021} for more details).   
We test the accuracy of our code at our standard resolution for LS220 EoS case with mass of $1.4 M_{\odot}$. Applying the fast Fourier transform on the density oscillations at the neutron star's centre to extract modes frequencies, we find our numerical results close enough to the $f$-mode and the first harmonic $H1$ derived from the perturbation equations. For $f$-mode we get $4.059$ kHz from simulation and $4.065$ kHz from the perturbation equations and for $H1$ $6.959$ kHz from simulation and $7.043$ kHz from perturbation equations. Fig.~\ref{fig:ls220_cowling} illustrates the frequency spectrum from our numerical simulation compared with the results from the analytical solution of the perturbation equations.

\subsection{$f$-mode frequency at different resolutions}
\label{sec:dd2_conv}

\begin{figure}
    \centering
    \includegraphics[width=0.49\textwidth]{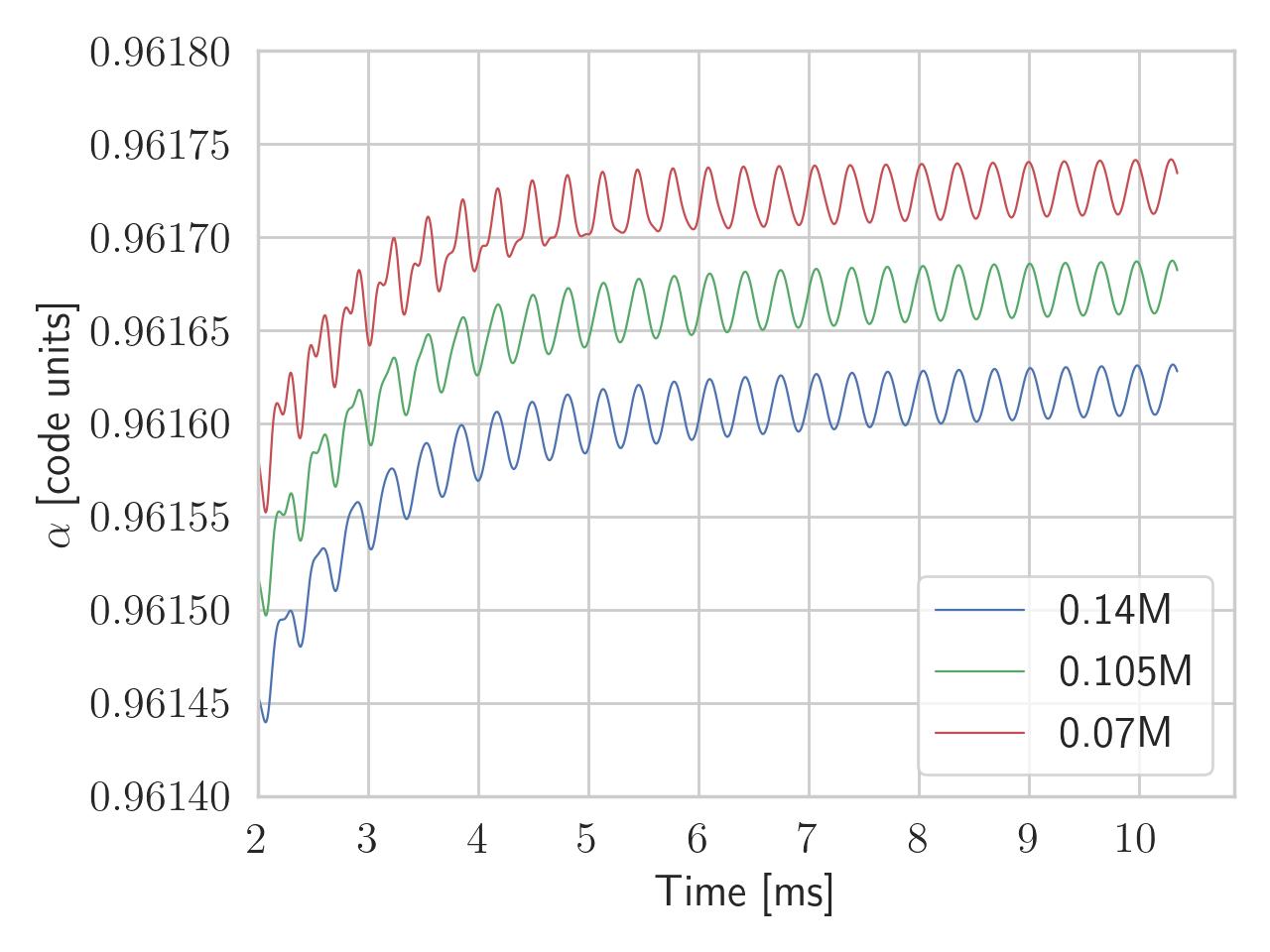}
    \includegraphics[width=0.49\textwidth]{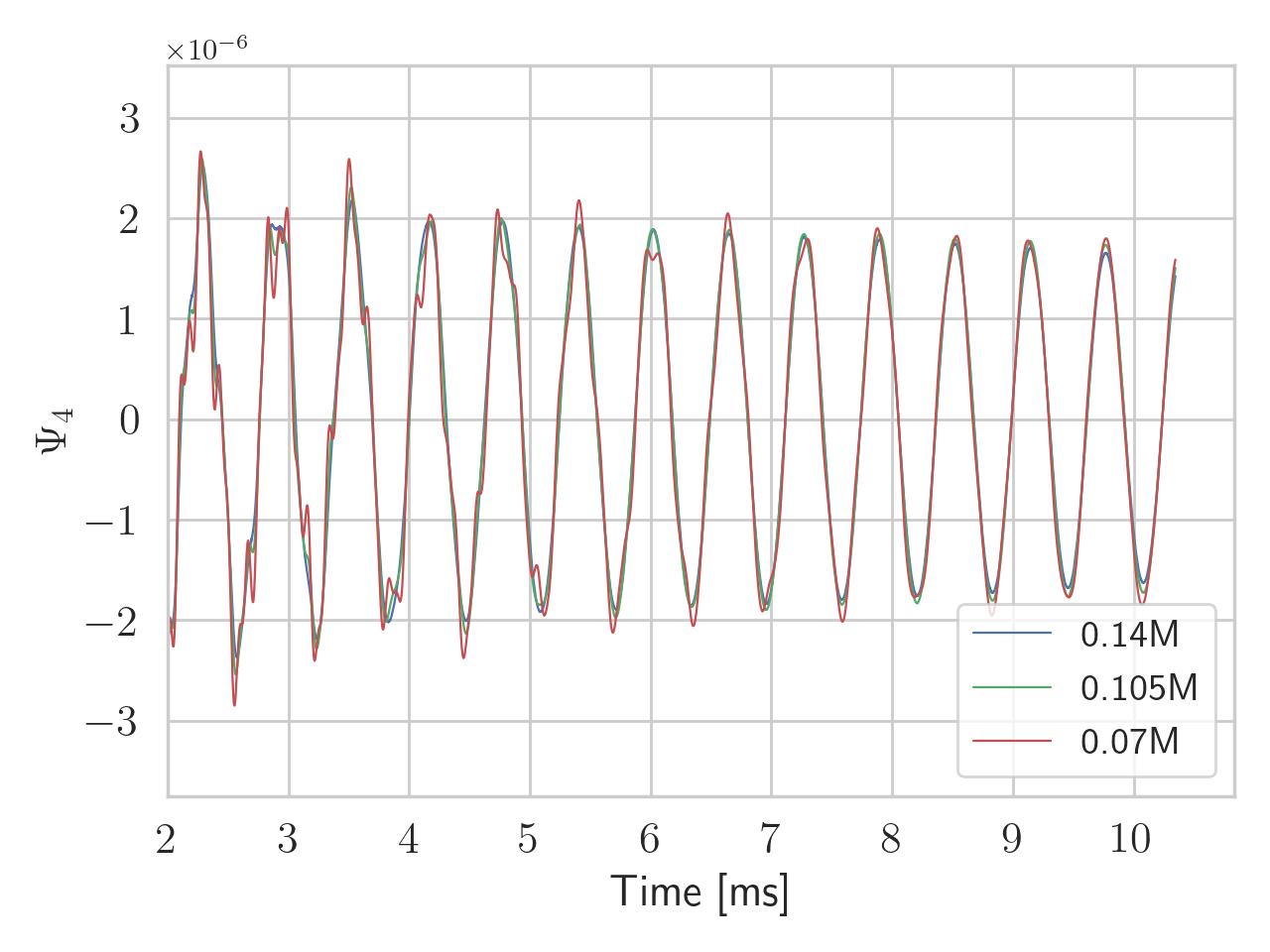}
    \includegraphics[width=0.49\textwidth]{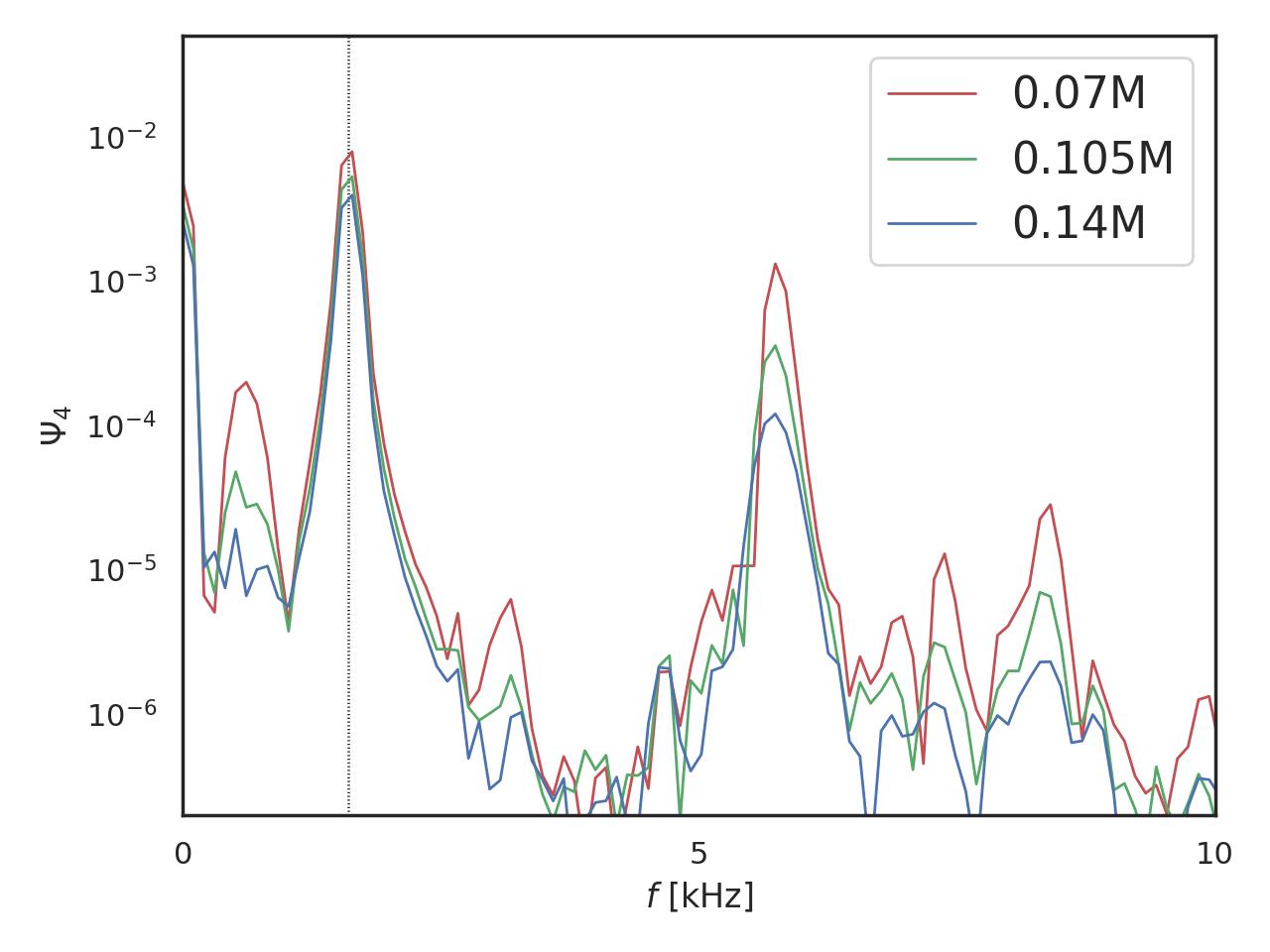}
    \includegraphics[width=0.49\textwidth]{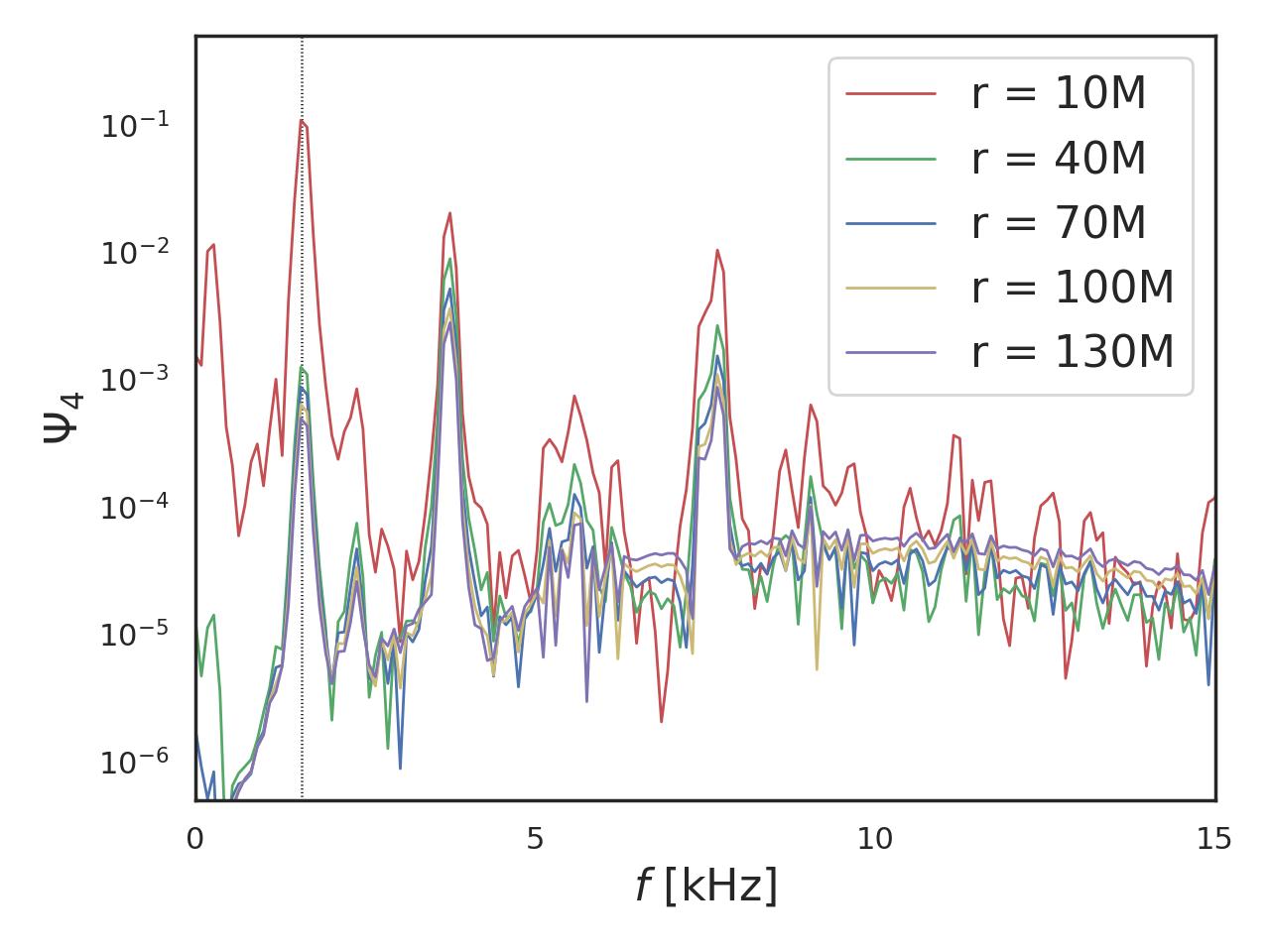}
    \caption{The central lapse for three different resolutions (top left); $\Psi_4$ plotted at three different resolution. The change in resolution does not change the $\Psi_4$ data (top right). The FFT of $\Psi_4$ at three different resolution shows the $f$-mode value does not change. The black dashed line represents the value of $f$-mode frequency (bottom left). The FFT of $\Psi_4$ at 5 different radii with a refinement layer between each (bottom right).
    }
    \label{fig:convergence_test}
\end{figure}

The performance of \texttt{WhiskyTHC} code has been tested before~\cite{thc1}. Here we only test the accuracy of our results within the selected resolution for our simulations. 
In order to claim that the resolution is in the convergence regime, we perform a multiple resolutions test for a single case, i.e. DD2 EoS with $M=1.4 M_{\odot}$. 
We choose our standard resolution with grid spacing equals to $0.105M$ as the intermediate level, and $0.07M$ and $0.14M$ as the higher and lower resolutions respectively. The $\Psi_4$ data extracted at $10M$ for this test.

The results of these tests are presented in Fig.~\ref{fig:convergence_test}. The top-left panel shows the lapse function varying with time in the simulation. One can see that increasing the resolution result in the convergence toward one solution. On the top-right panel, we present $\Psi_4$ for different resolutions. This plot shows that increasing the resolution does not change the gravitational wave significantly, but it reduces the noise as expected. The bottom-left panel shows the FFT of $\Psi_4$. This figure shows that the changes in frequency are negligible by varying the resolution, which confirms the accuracy of our $f$-mode frequency measurements.
These results indicate that the observed quantities converge to one solution as we move from low to high resolutions, and the value of $f$-mode frequencies do not change significantly across different resolutions.
This convergence study confirms that the intermediate resolution used in our numerical simulations, to report the $f$-mode frequencies, is in the convergence regime.

\subsection{$f$-mode frequency at different $\Psi_4$ extraction radii}

As the last numerical test, we investigate the accuracy of the $\Psi_4$ functions extracted from different radii. For this test we choose our polytropic case with $\Gamma=2$, $\kappa=100$ and $M=1.4M_{\odot}$, with minimum grid spacing equals to $0.06M$ for the resolution. The $\Psi_4$ outputs are extracted at $r=10,40,70,100,130M$ for the Fourier transform. 

In Fig.~\ref{fig:convergence_test} (bottom right), we show that the $f$-mode frequency does not change with different radii of extraction of $\Psi_4$ for Polytropic equation of state. We also notice that $f$-mode is most prominent and has less noise for $r=10M$, and hence we chose it for comparison in with other equations of state. 


\section{Analysis and Results}
\label{sec:results}

We evolve a set of non-rotating configurations in the mass range of $1.2-2.0 M_\odot$ for each of the EoS, described in Sec.~\ref{subsec:EoS} and test our obtained results against the established results from perturbative approach. The mass, radius, central density for these have been listed in the first four columns of Tab.~\ref{tab:$f$-modes}.

We also notice that change in resolution does not affect the extracted $f$-mode frequencies (Fig.~\ref{fig:convergence_test}). We run the simulations for $2100 M \approx 10.35 ms$.

\subsection{fundamental ($f$)-modes}
\label{subsec:fmodes}
   
\begin{figure*}
    \centering
    \includegraphics[width=0.45\textwidth]{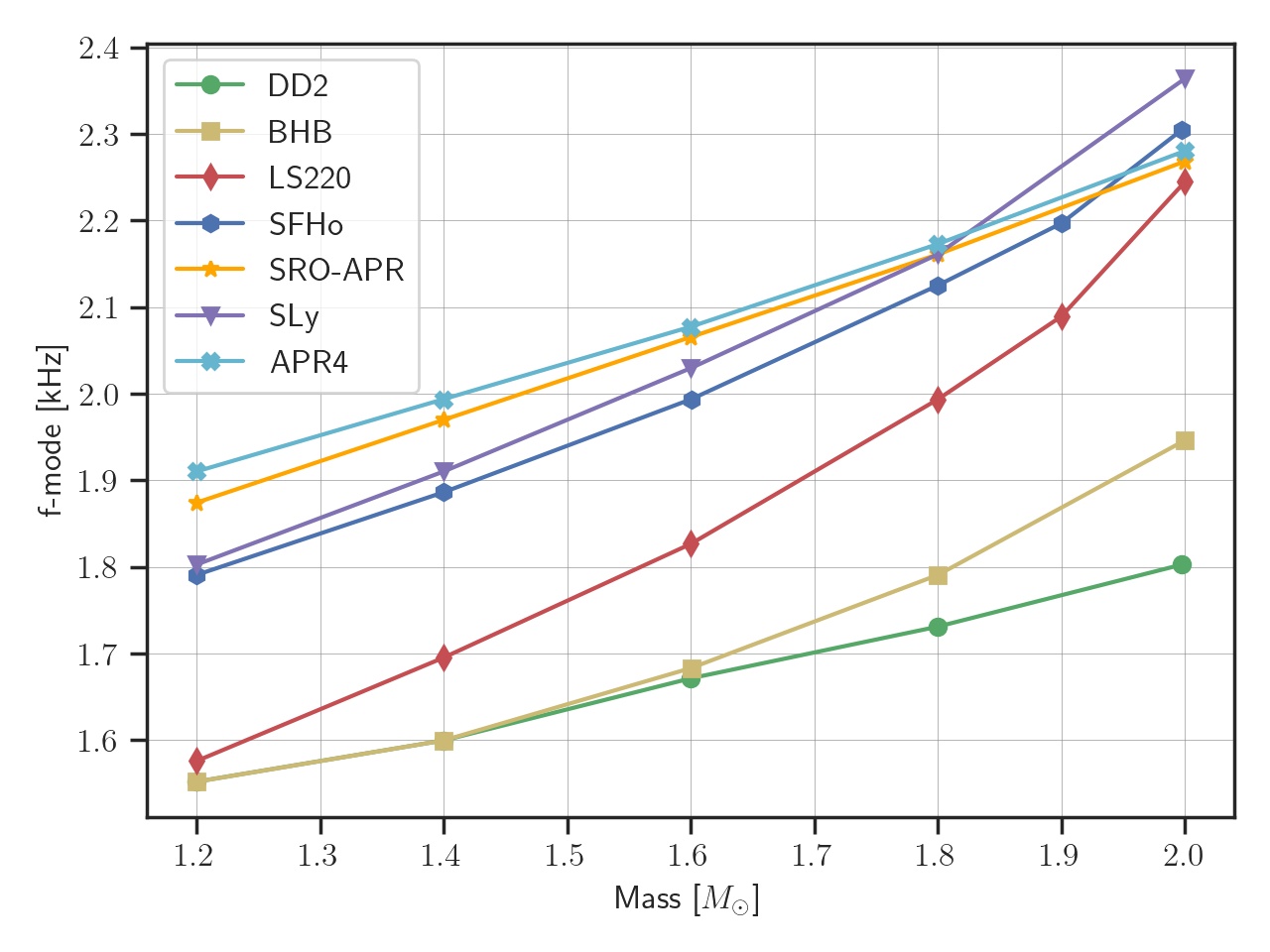}
    \includegraphics[width=0.45\textwidth]{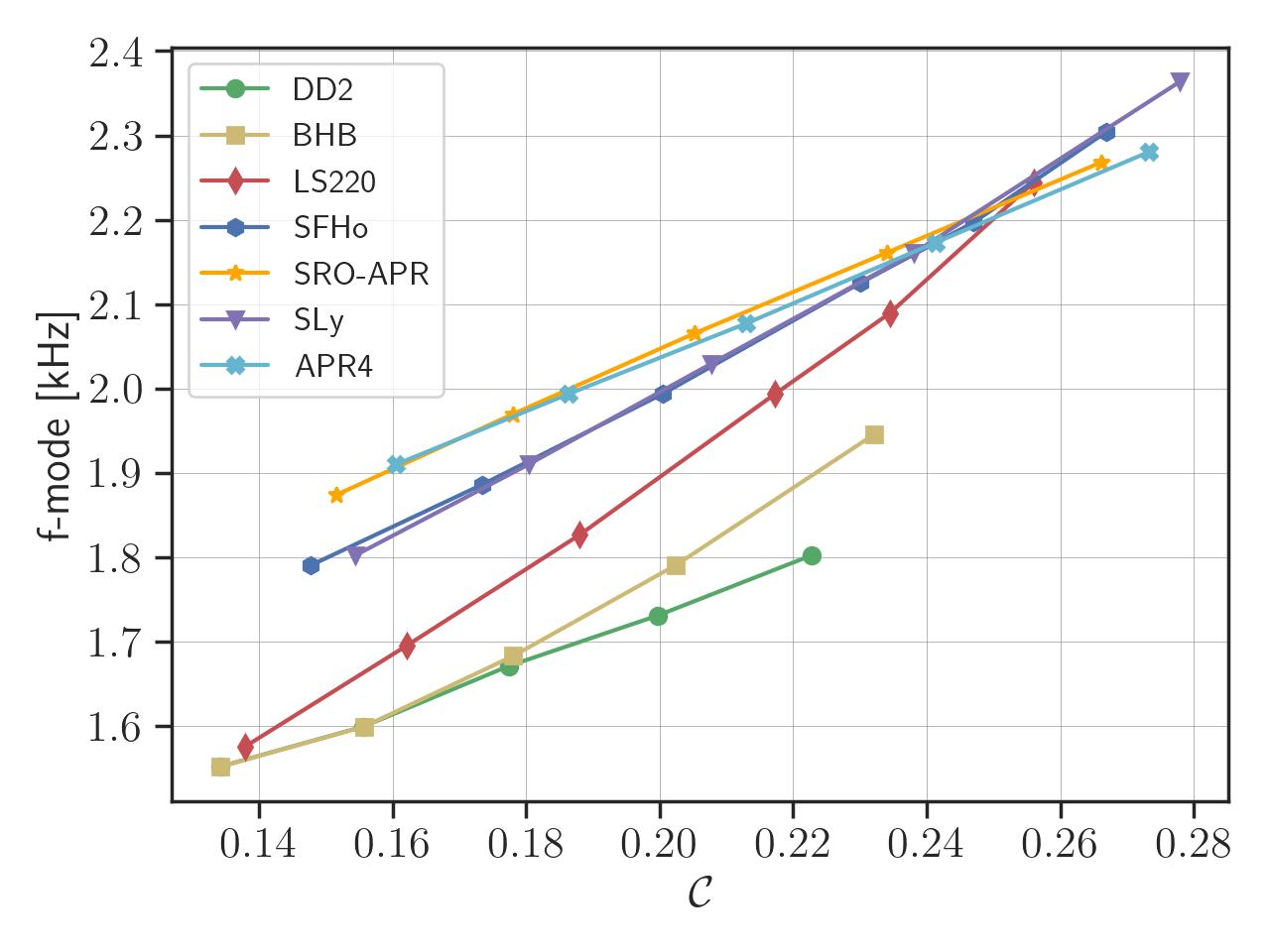}
    \includegraphics[width=0.45\textwidth]{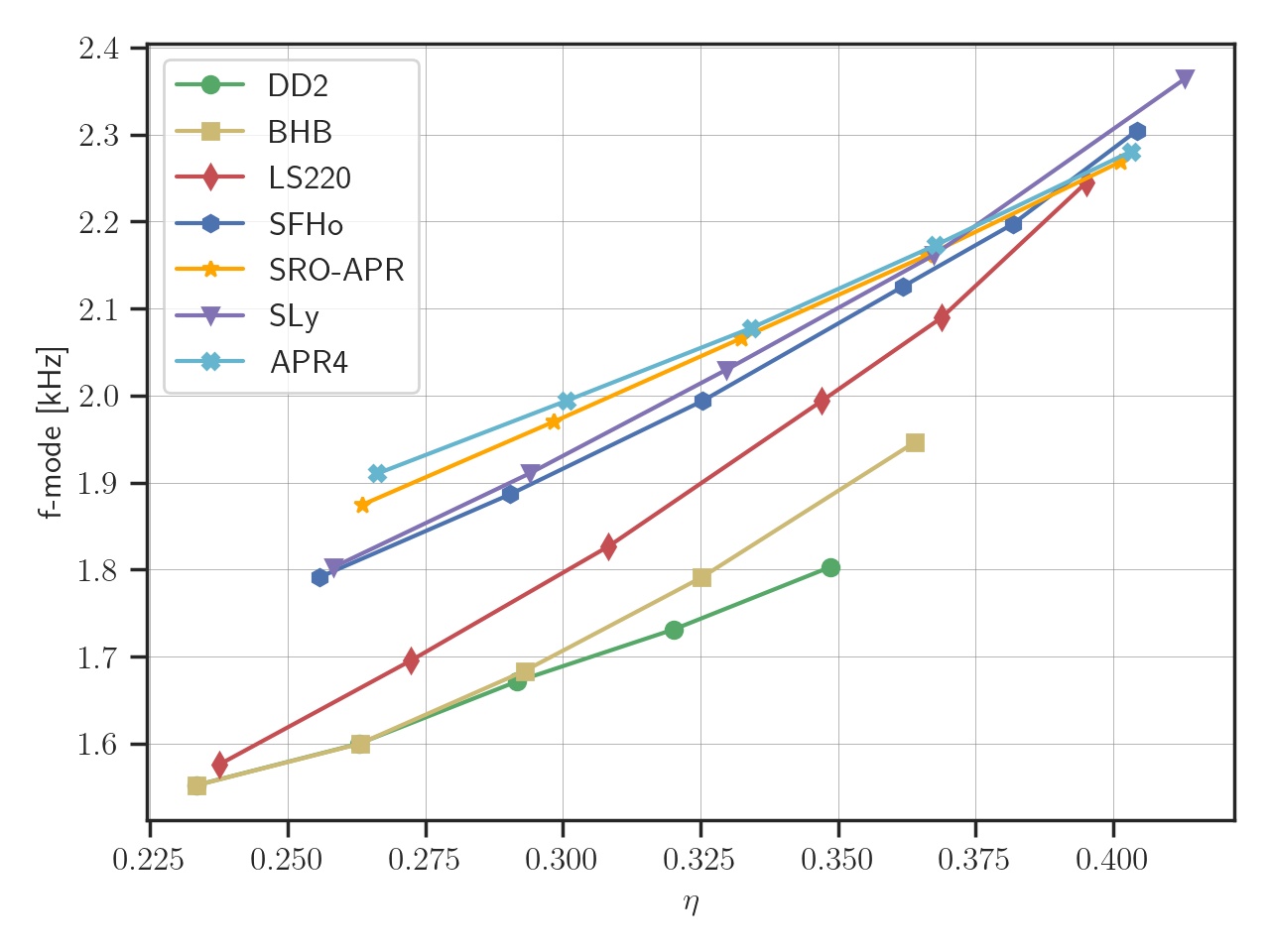}
    \includegraphics[width=0.45\textwidth]{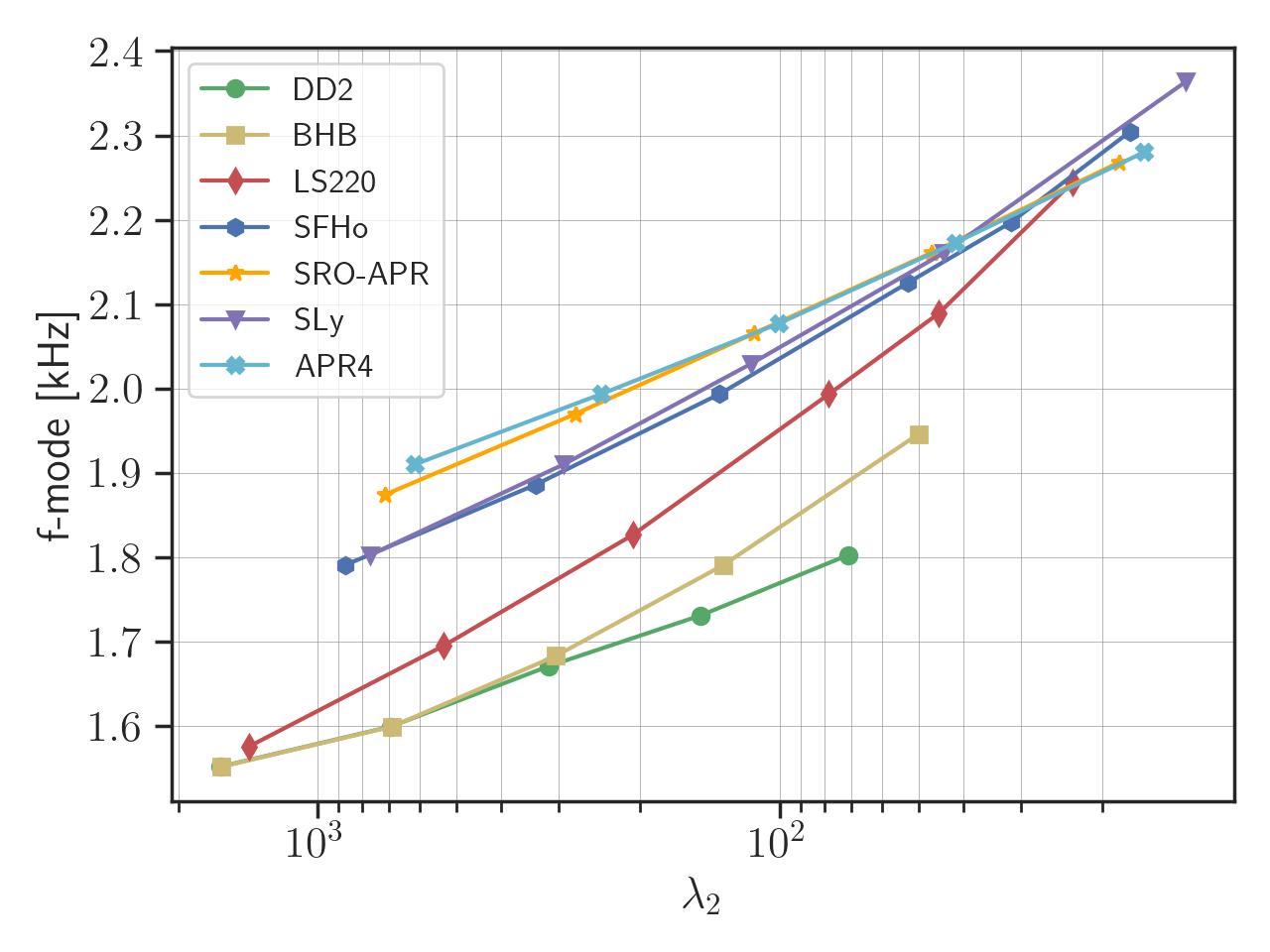}
    \caption{ $f$-mode frequency for different EoS is shown in different panels with respect to mass $M$ (top left); compactness $\mathcal{C} = M/R$, $R$ being the radius of the star (top right); effective compactness $\eta  = \sqrt{M^3/I}$, where $I$ is the moment of inertia  (bottom left) and tidal deformability $\lambda_2$ as in Eq.~\ref{eqn:tidal_deform} (bottom right). The softer EoS has larger $f$-mode frequencies as compared to the intermediate (LS220) and stiffer (DD2 and BHB) EoS.}
    \label{fig:freq_vs_param}
\end{figure*}

We present our models and results of our simulations in Tab.~\ref{tab:$f$-modes}. In Fig.~\ref{fig:freq_vs_param} 
(top panels) we plot the $f$-mode frequency value in terms of mass and compactness for the considered set of equations of state. 

We notice that the $f$-mode frequency is much smaller for the stiff EoS such as DD2 and BHB. It increases for the higher masses and softer EoS. Based on the frequency band of the observed gravitational signal and the inferred mass, it is possible to put bounds on the softness/stiffness of the EoS that characterises the interior of the neutron star. For example, if the detected $f$-mode frequency is below $1.8$ kHz for a star having mass above $1.4 M_\odot$, then all the considered soft EoS would be ruled out. On the other hand, frequency above $1.8$ kHz would permit only some stiff EoS if the object is more massive i.e. $\approx 1.8 M_\odot$ or more and compactness above $0.20$. Similar trends we see in terms of effective compactness $\eta$ and tidal deformability $\lambda_2 $ (see bottom panels of Fig.~\ref{fig:freq_vs_param}). We also observe that while few of the considered EoS show linear trend, some of the other, such as BHB, LS220, SLy and SFHo deviate and show a faster rise in $f$-mode values for larger parameter values (as could be seen in all the four panels of Fig.~\ref{fig:freq_vs_param}). This doesn't seem to be dependent only on the softness or stiffness of the matter, but could be due to the finer micro-physics involved. To understand this better, in our follow-up studies, we plan to consider a larger set of EoS, including the ones having hyperons, strange quark matter, etc.

\begin{figure}
    \centering
    \includegraphics[width=0.55\textwidth]{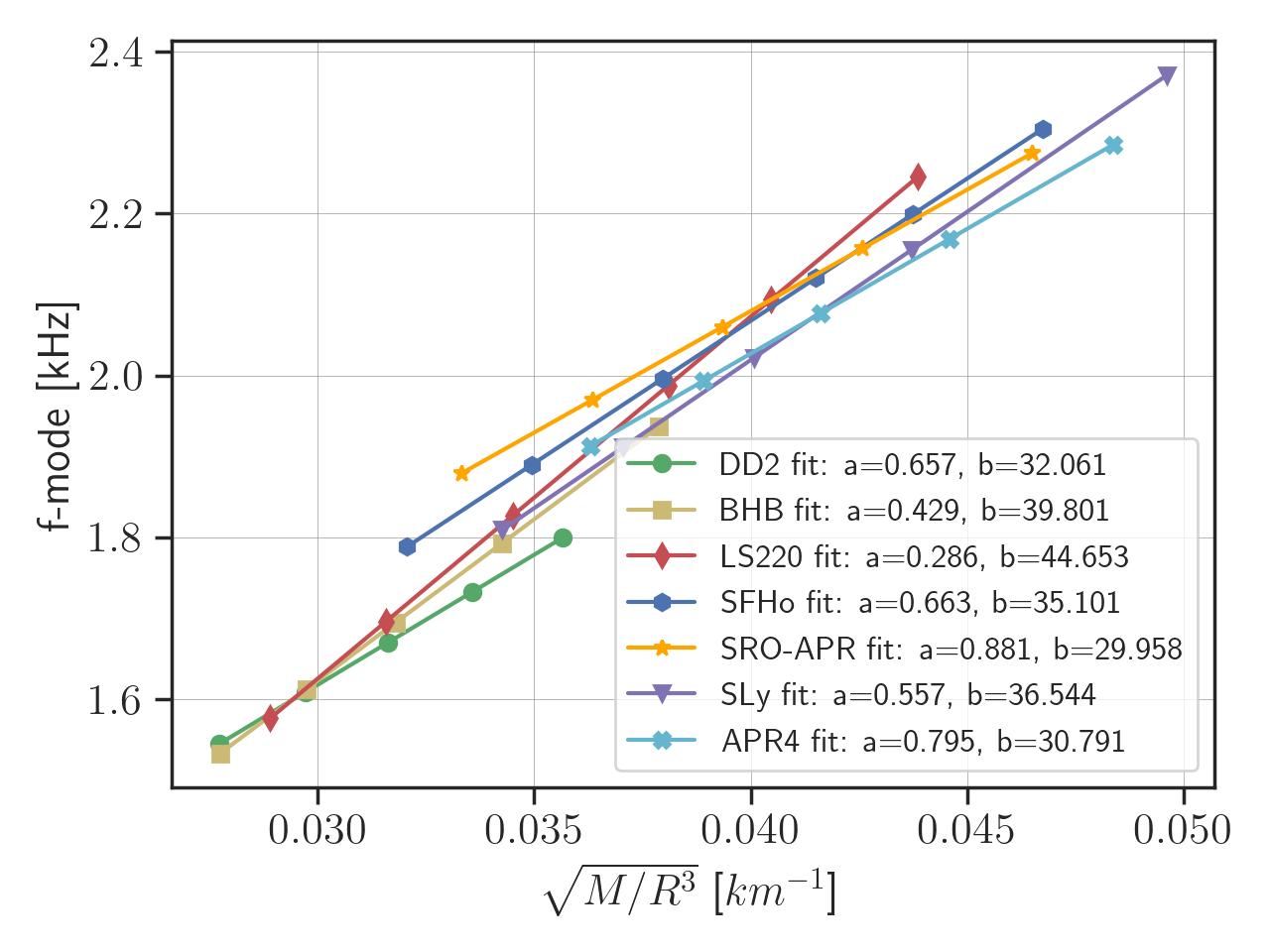}
    \caption{Fit from Eq.~\ref{eqn:chirenti_fit}. The $a$ and $b$ values are listed on the figure above and also Tab.~\ref{tab:chirenti_fit}.}
    \label{fig:chirenti_fit}
\end{figure}

In Fig.~\ref{fig:chirenti_fit} we present the fits of our models with the relation as given in Ref.~\cite{Chirenti:2015PRD} :
\begin{equation}
\label{eqn:chirenti_fit}
    f = a + b ~\sqrt{\frac{M}{R^3}}
\end{equation}

We find a good linear fit for our data across the EoS used for our study and list the values of $a$ and $b$ in Tab.~\ref{tab:chirenti_fit}. LS220 EoS shows the steepest change (the red line), followed by BHB (the Beige color line) in Fig.~\ref{fig:chirenti_fit}.
\begin{table}
    \centering
    {\small
    \begin{tabular}{|c|c|c|}
    \hline
        EoS & a & b \\
    \hline
        DD2 & 0.657 & 32.061 \\
        BHB & 0.429 & 39.801 \\
        LS220 & 0.286 & 44.653 \\
        SFHo & 0.663 & 35.101 \\
        SRO-APR & 0.881 & 29.958 \\
        SLy & 0.557 & 36.544 \\
        APR4 & 0.795 & 30.791 \\
    \hline
    \end{tabular}
    }
    \caption{Data fitted for the various compact star models with Eq.~\ref{eqn:chirenti_fit}. We find that our results are in agreement with the Ref.~\cite{Chirenti:2015PRD}.}
    \label{tab:chirenti_fit}
\end{table}
We compare our fit obtained in Tab.~\ref{tab:chirenti_fit} to the Table II in Ref.~\cite{Chirenti:2015PRD} for the LS220 and APR4 equation of states, and find our results to be in agreement (within $10\%$).

\subsection{Universal Relations}
\label{subsec:ur}
In order to study, whether the equations of state specific trends that we notice above, contribute to the deviation from the established Universal relations or not, we verify some of the URs. First, we compute the relation between $f$-mode frequency and the tidal deformability as given in Ref.~\cite{hinderer_nature,ChanUR:PhysRevD.90.124023}\\
\begin{equation}
\label{eqn:UR_f-l}
    M \omega = \sum_i a_i ~(\xi)^i,
\end{equation}
\\
where $a_i$ are the numerical coefficients presented in Tab.~\ref{tab:f-Love_fit}. $M$ is the mass of the star and $\omega$ is the angular $f$-mode frequency. $\xi = log(\lambda_2)$, where $\lambda_2$ is the dimensionless electric
tidal deformability calculated from the tidal Love number $k_2$ as  \cite{ChanUR:PhysRevD.90.124023}:
\begin{equation}
\label{eqn:tidal_deform}
    \lambda_2 = \frac{2}{3} \frac{k_2}{(M/R)^{5}}
\end{equation}

The tidal Love number $k_2$ is calculated by solving the metric perturbation equation as defined in Ref.~\cite{Hinderer_2008}.
For this computation we integrate Eq. (15) from~\cite{Hinderer_2008} for the metric perturbation function $H$, from center to surface using fourth order Runge-Kutta method. We use the Runge-Kutta ODE solver with adaptive step size routine from Numerical recipe~\cite{NumRec}.
Finally, the tidal Love number $k_2$ is computed from Eq.(23) from~\cite{Hinderer_2008}. Figure~\ref{fig:freq_vs_param} (bottom-right panel) shows the relation of tidal deformability $\lambda_2$ with the $f$-mode.

\begin{figure}
    \centering
    \includegraphics[width=0.55\textwidth]{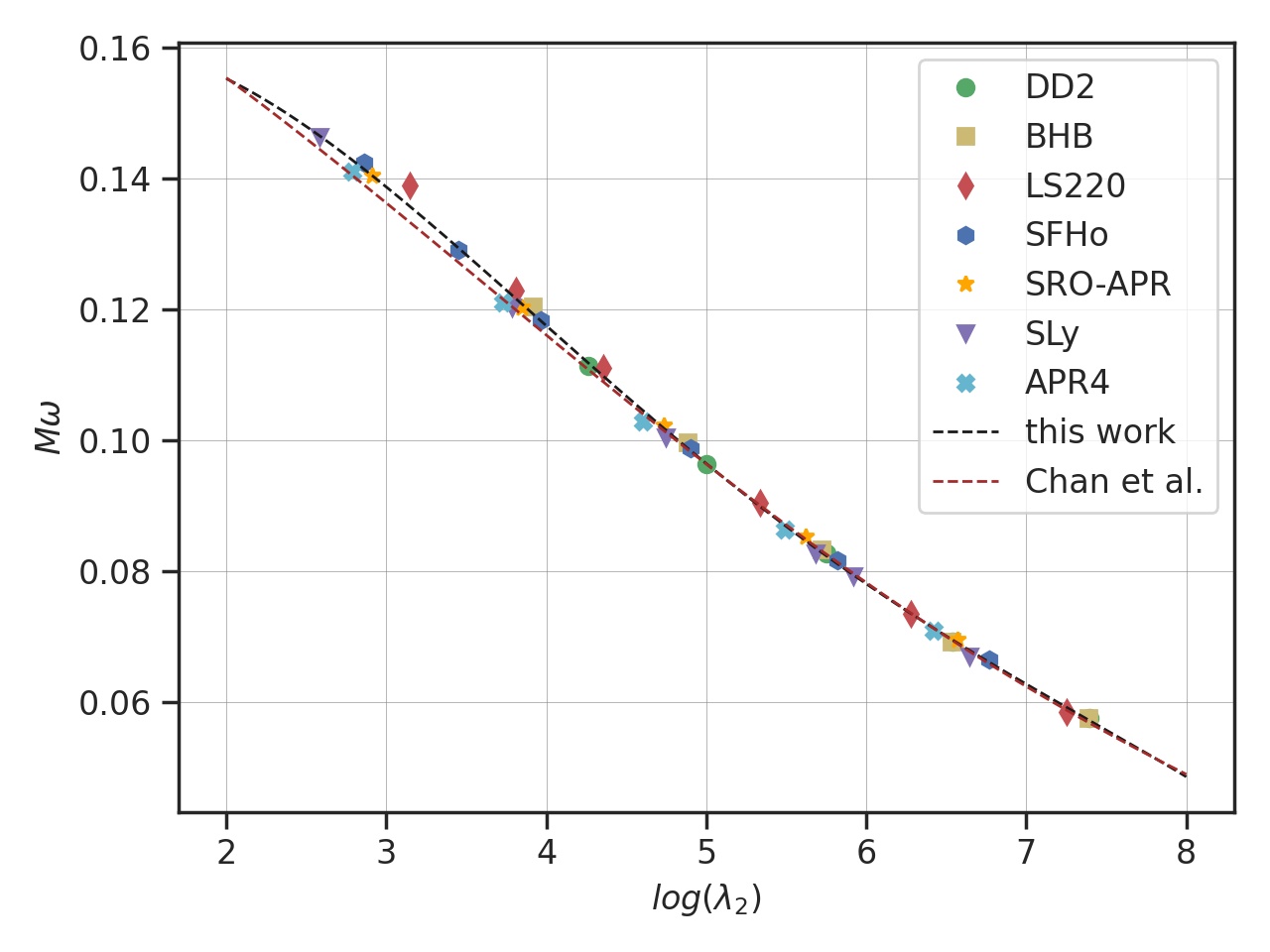}
    \caption{The universality observed between tidal deformability and $f$-modes. Here, we compare our fit with that of Ref.~\cite{ChanUR:PhysRevD.90.124023}.}
    \label{fig:f-love_UR}
\end{figure}

The comparison and deviation for  $f$-Love UR, Eq.~\ref{eqn:UR_f-l}
is carried out in two ways: first, using these equations we compute the coefficients for our data. Second, using the same values of coefficients as given in Ref.~\cite{ChanUR:PhysRevD.90.124023}
but, with the $f$, $\lambda_2$ and $\eta$ that we calculate for each of our simulation. 
The last three columns of Tab.~\ref{tab:$f$-modes} show comparison and percent deviations with $f$-Love UR's. Table~\ref{tab:f-Love_fit} lists the coefficients for Ref.~\cite{ChanUR:PhysRevD.90.124023} and the ones computed for our data. In Fig.~\ref{fig:f-love_UR} we compare the our results with that of Ref.~\cite{ChanUR:PhysRevD.90.124023}. 

\begin{figure}
    \centering
    \includegraphics[width=0.55\textwidth]{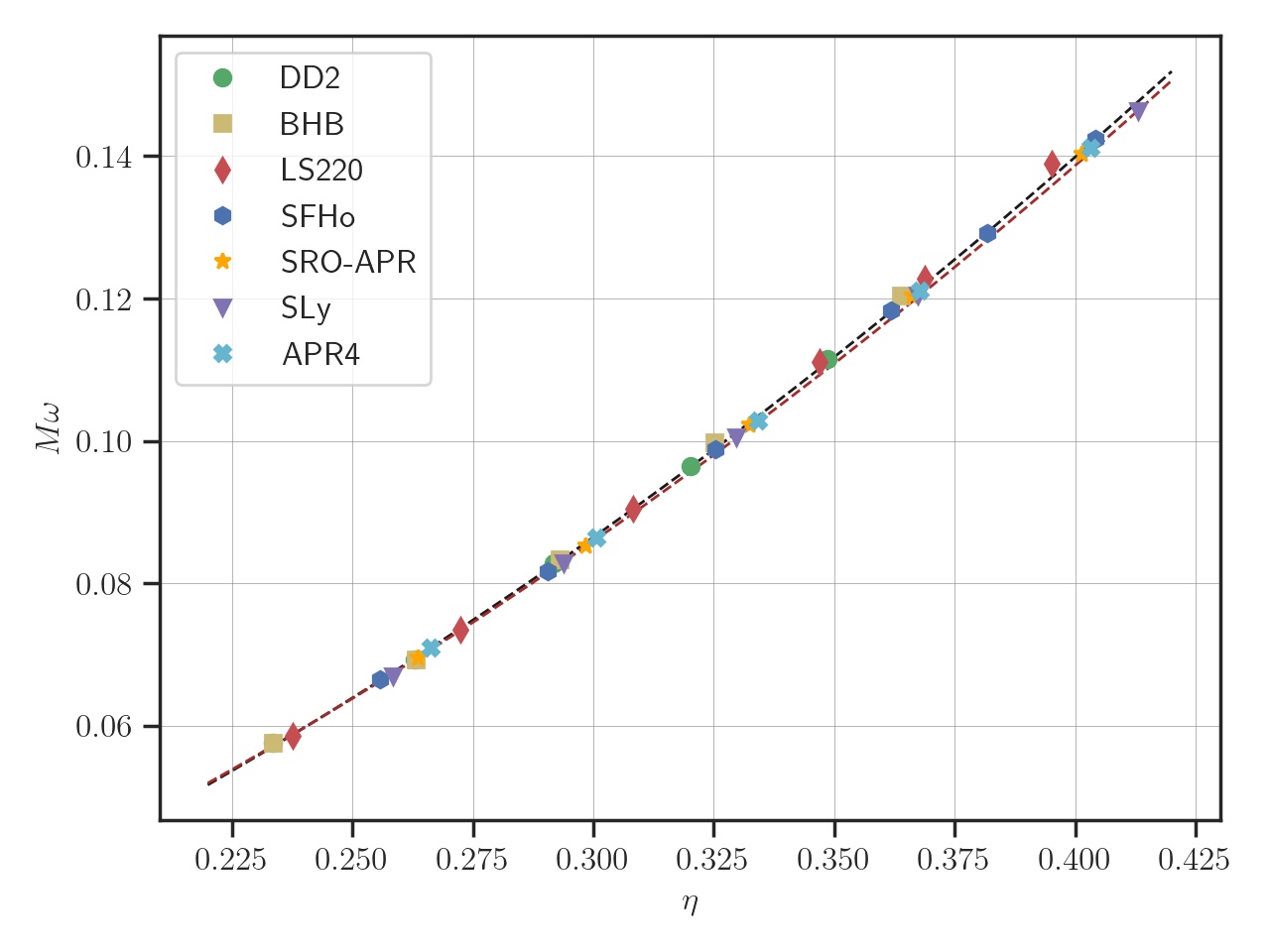}
    \caption{The universality observed between $\eta$ and $f$-mode. The black dashed line is the fit that we obtained for Eq.~\ref{eqn:UR_eta-f}. The brown dashed line is the one presented in the Ref.~\cite{Chirenti:2015PRD}.}
    \label{fig:eta_v_mw}
\end{figure}

It has been observed that universal behaviour also exist between the $f$-mode and the
effective compactness $\eta$ \cite{Lau:2010, Chirenti:2015PRD}. We also test these URs described by the model
\begin{equation}
\label{eqn:UR_eta-f}
    M \omega = c_1 + c_2~\eta + c_3~\eta^2,
\end{equation}
\\
where $\eta = \sqrt{M^3/I}$ and $I$ is the moment of inertia. We obtain the fit as $c_1 = -0.00747$, $c_2 = 0.1471$ and $c_3 = 0.55328$. We show this universality in the Fig.~\ref{fig:eta_v_mw}. Our results agree well with the results of the earlier works \cite{Lau:2010, Chirenti:2015PRD}.
The fundamental mode $f$-mode and effective compactness $\eta$ that we compute for our configurations are also plotted in Fig~\ref{fig:freq_vs_param} (bottom left panel).\\

Universal relation between the compactness and tidal Love numbers have been discussed in ~\ref{appendixA}.

\begin{table}
    \centering
    {\small
    \begin{tabular}{|c|c|c|c|c|c|c|}
        \hline
         EoS & \shortstack{Mass \\ ($M_{\odot}$)} & \shortstack{Radius \\ ($km$)} & \shortstack{Central Density \\ ($g/cm^3$)} & \shortstack{$f$-mode \\ ($kHz$)} & k2 &
         \shortstack{(\%)-difference \\ Chan et al.~\cite{ChanUR:PhysRevD.90.124023} \\ v. This work} \\
         \hline
		 Polytrope & 1.35 & 13.80 & $8.22 \times 10^{14}$ & 1.6237 & 0.0805 & 0.22 \\
		  & 1.4 & 14.15 & $7.91 \times 10^{14}$ & 1.5730 & 0.0793 & 0.26 \\
		 \hline
		  & 1.2 & 13.20 & $5.25 \times 10^{14}$ & 1.5522 & 0.1061 & 0.25 \\
		  & 1.4 & 13.27 & $5.73 \times 10^{14}$ & 1.5998 & 0.0953 & 0.33 \\
		 DD2 & 1.6 & 13.31 & $6.27 \times 10^{14}$ & 1.6715 & 0.0833 & 0.53 \\
		  & 1.8 & 13.30 & $6.9 \times 10^{14}$ & 1.7312 & 0.0709 & 0.05 \\
		  & 2.0 & 13.24 & $7.67 \times 10^{14}$ & 1.8028 & 0.0586 & 0.86 \\
		 \hline
		  & 1.2 & 13.20 & $5.25 \times 10^{14}$ & 1.5521 & 0.1058 & 0.24 \\
		  & 1.4 & 13.27 & $5.84 \times 10^{14}$ & 1.5998 & 0.0947 & 0.34 \\
		 BHB & 1.6 & 13.26 & $6.73 \times 10^{14}$ & 1.6834 & 0.0821 & 0.53 \\
		  & 1.8 & 13.13 & $8.0 \times 10^{14}$ & 1.7909 & 0.0675 & 0.08 \\
		  & 2.0 & 12.71 & $1.02 \times 10^{15}$ & 1.9461 & 0.0508 & 1.29 \\
		 \hline
		  & 1.2 & 12.84 & $6.16 \times 10^{14}$ & 1.5760 & 0.1056 & 0.14 \\
		  & 1.4 & 12.75 & $7.19 \times 10^{14}$ & 1.6954 & 0.0898 & 0.46 \\
		 LS220 & 1.6 & 12.56 & $8.47 \times 10^{14}$ & 1.8267 & 0.0734 & 0.35 \\
		  & 1.8 & 12.23 & $1.03 \times 10^{15}$ & 1.9936 & 0.0568 & 0.74 \\
		  & 1.9 & 11.96 & $1.16 \times 10^{15}$ & 2.0893 & 0.0482 & 1.41 \\
		  & 2.0 & 11.53 & $1.37 \times 10^{15}$ & 2.2446 & 0.0383 & 1.80 \\
		 \hline
		  & 1.2 & 11.99 & $7.41 \times 10^{14}$ & 1.7909 & 0.0922 & 0.21 \\
		  & 1.4 & 11.92 & $8.4 \times 10^{14}$ & 1.8864 & 0.0793 & 0.54 \\
		 SFHo & 1.6 & 11.79 & $9.6 \times 10^{14}$ & 1.9938 & 0.0653 & 0.06 \\
		  & 1.8 & 11.56 & $1.12 \times 10^{15}$ & 2.1252 & 0.0511 & 1.23 \\
		  & 1.9 & 11.36 & $1.24 \times 10^{15}$ & 2.1968 & 0.0436 & 1.71 \\
		  & 2.0 & 11.05 & $1.43 \times 10^{15}$ & 2.3042 & 0.0355 & 1.72 \\
		 \hline
		  & 1.2 & 11.68 & $8.03 \times 10^{14}$ & 1.8744 & 0.0862 & 0.32 \\
		  & 1.4 & 11.61 & $8.9 \times 10^{14}$ & 1.9699 & 0.0745 & 0.50 \\
		 SRO-APR & 1.6 & 11.51 & $9.87 \times 10^{14}$ & 2.0655 & 0.0620 & 0.26 \\
		  & 1.8 & 11.35 & $1.11 \times 10^{15}$ & 2.1610 & 0.0495 & 1.37 \\
		  & 2.0 & 11.09 & $1.27 \times 10^{15}$ & 2.2684 & 0.0368 & 1.75 \\
		 \hline
		  & 1.2 & 11.47 & $7.79 \times 10^{14}$ & 1.8028 & 0.1016 & 0.28 \\
		  & 1.35 & 11.46 & $8.58 \times 10^{14}$ & 1.8983 & 0.0885 & 0.55 \\
		  & 1.4 & 11.46 & $8.86 \times 10^{14}$ & 1.9103 & 0.0840 & 0.52 \\
		 SLy & 1.6 & 11.37 & $1.02 \times 10^{15}$ & 2.0296 & 0.0670 & 0.24 \\
		  & 1.8 & 11.16 & $1.2 \times 10^{15}$ & 2.1610 & 0.0506 & 1.43 \\
		  & 2.0 & 10.62 & $1.55 \times 10^{15}$ & 2.3639 & 0.0328 & 1.41 \\
		 \hline
		  & 1.2 & 11.04 & $8.24 \times 10^{14}$ & 1.9103 & 0.0984 & 0.40 \\
		  & 1.4 & 11.09 & $9.12 \times 10^{14}$ & 1.9938 & 0.0818 & 0.44 \\
		 APR4 & 1.6 & 11.09 & $1.01 \times 10^{15}$ & 2.0774 & 0.0661 & 0.41 \\
		  & 1.8 & 11.01 & $1.14 \times 10^{15}$ & 2.1729 & 0.0512 & 1.49 \\
		  & 2.0 & 10.81 & $1.32 \times 10^{15}$ & 2.2804 & 0.0371 & 1.66 \\
		 \hline
    \end{tabular}
    }
    \caption{$f$-modes for the different EoS for different masses. A clear trend is visible. Frequency increases with mass and softness of the EoS.
    The last column lists the difference between the UR provided in Ref.~\cite{ChanUR:PhysRevD.90.124023} and our work.}
    \label{tab:$f$-modes}
\end{table}

\begin{table}
    \centering
    {\small
    \begin{tabular}{|c|c|c|c|c|c|}
        \hline
         Fit & $a_0$ & $a_1$ & $a_2$ & $a_3$ & $a_4$ \\
         \hline
         Chan et al. \cite{ChanUR:PhysRevD.90.124023} & $1.820 \times 10^{-1}$ & $-6.836 \times 10^{-3}$ & $-4.196 \times 10^{-3}$ & $5.215 \times 10^{-4}$ & $-1.857 \times 10^{-5}$  \\
         This work & $1.442 \times 10^{-1}$ & $3.005 \times 10^{-2}$ & $-1.607 \times 10^{-2}$ & $2.092 \times 10^{-3}$ & $-9.247 \times 10^{-5}$  \\
         \hline
    \end{tabular}
    }
    \caption{Fitting parameters $a_i$ from Eq.~\ref{eqn:UR_f-l} for the $f$-Love universal relations, comparing this work and Ref.~\cite{ChanUR:PhysRevD.90.124023}. The polytropic models are excluded in $a_i$ values calculations.}
    \label{tab:f-Love_fit}
\end{table}

\subsection{Calculation of damping times}
\label{subsec:dampingtimes}

\begin{figure}
    \centering
    \includegraphics[width=0.55\textwidth]{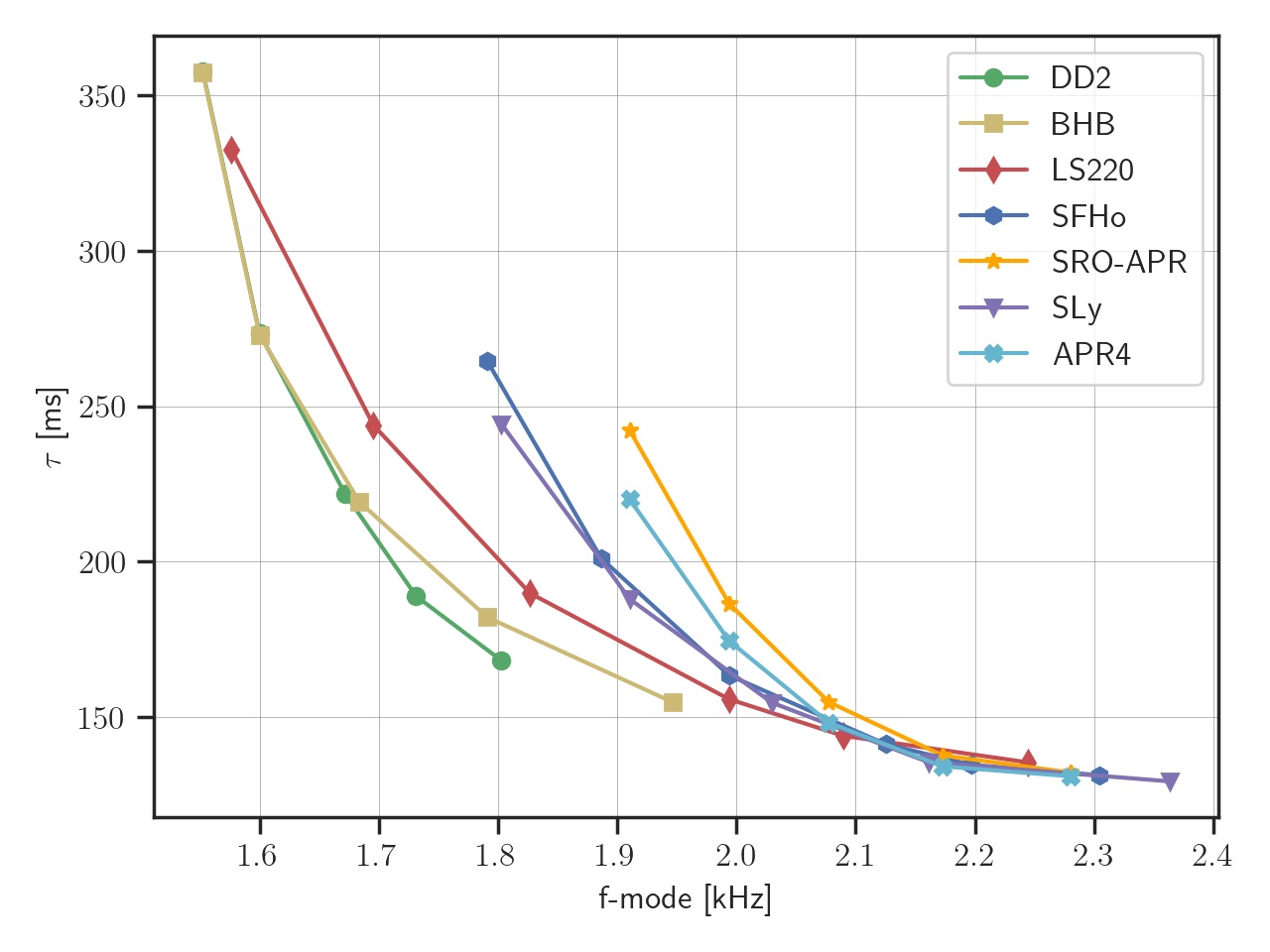}
    \caption{$f$-mode vs the damping time for all the models. The damping time here is calculated as $(\tau_1 + \tau_2)/2$ from Eq.~\ref{eqn:damping_time_MR} and Eq.~\ref{eqn:damping_time_I}.}
    \label{fig:fvtau}
\end{figure}

We compute fundamental frequency from $\Psi_4$ data of our simulations (see discussion in the beginning of Sec.~\ref{sec:results}) by evolving each initial configuration for about $2100 M$ ($ \approx 10 $ ms). This duration is insufficient, and it is required to have a longer simulation and higher resolution to extract reliable damping times. Thus, we choose the recently established relations, which use compactness and effective compactness for calculating the damping times \cite{damping_nick_2018, Chirenti:2015PRD}:

\begin{equation}
\label{eqn:damping_time_MR}
    \frac{M}{\tau_1} = 0.112 \left(\frac{M}{R}\right)^4 - 0.53 \left(\frac{M}{R}\right)^5 + 0.628 \left(\frac{M}{R}\right)^6,
\end{equation}

\begin{equation}
\label{eqn:damping_time_I}
    \frac{I^2}{M^5 \tau_2} = 0.0068 - 0.025 ~\eta^2,
\end{equation}
where $\tau_1$ and $\tau_2$ are the damping times in terms of compactness $\mathcal{C} = \frac{M}{R}$ and effective compactness $\eta$, respectively . 
Figure~\ref{fig:fvtau} shows damping time vs frequency plot, where the damping time is taken to be the average of $\tau_1$ and $\tau_2$. We find that for the polytrope case one of the relations overestimates the damping time obtained from linear perturbations methods \cite{baiotti_pert, rosofsky} while the other underestimates, taking an average of $\tau_1$ and $\tau_2$ gives us a value close to the expected value for the polytrope, hence we report the average values. In the subsequent study, we intend to extract damping times by evolving our systems for a much longer time and at a much higher resolution as one needs to be sure that the damping obtained is not due to numerical errors.

\subsection{Comparison with compact binary merger simulations}

In the context of a binary system, the tidal force comes from a companion star in circular motion. The spherical harmonic expansion of the full tidal potential is given by Eq.(2.2) from Lai (1994)~\cite{Lai1994}:

\begin{equation}
 U = -GM \sum_{l,m} W_{lm}\frac{r^l}{A(t)^{l+1}} Y_{lm}(\theta, \phi ) e^{-im \Omega t},
 \label{eq:tide-expan}
\end{equation}

where $G$ is the gravitational constant, $M$ is the mass of the companion star, the coefficients $W_{lm}$ depend on $(l,m)$ with equation (2.3) from~\cite{Lai1994}, $A(t)$ is the binary separation and $\Omega$ is the orbital frequency. 

The initially seeded perturbation with $Y_{22}$ harmonics given by Eq.(\ref{eqn:perturbation}), corresponds to the leading quadrupole term of the spherical harmonic expansion of the tidal potential in Eq.(\ref{eq:tide-expan}).
Therefore, the initial setup of our single neutron star simulation helps us to artificially mimic the perturbation caused by the tidal field of the companion star during inspiral phase up to an acceptable order of accuracy. 

As a comparison we simulate couple of configuration to compare the $f$-mode frequency with Ref.~\cite{steinhoff2021spin}, where they extract the $f$-mode frequency using numerical relativity simulation of binary systems. We obtain an excellent match with their results, comparing to their black hole - neutron star binary simulation using $1.35M_{\odot}$ polytrope, we find a deviation of $0.6 \%$ and with their neutron star - neutron star binary simulation using $1.35_{\odot}M$ SLy EoS we see a deviation of only $0.03\%$. We further note that even by interpolating the mass vs $f$-mode data based on our simulations to get the $f$-mode frequency at $1.35 M_{\odot}$, the value which we get differs only by $0.9\%$ for the SLy EoS from the one reported in Ref.~\cite{steinhoff2021spin}.


\section{Conclusions}
\label{sec:conclusions}

In this work, we evolve isolated non-spinning neutron star in fully dynamical spacetime with nuclear EoS. 
We investigate the accuracy of our numerical methods by conducting several tests including the convergence, and Cowling evolution described in Sec.~\ref{sec:num_tests}. These tests confirm that our selected grid resolution is in the convergence regime, and their results match very well with the linear perturbation theory's solution for $f$-mode frequency measurement in the Cowling approximation.

For the main simulations using full general relativistic hydrodynamic evolution,
we consider a set of realistic EoS as listed in Sec.~\ref{subsec:EoS} to describe internal composition of the star. For each of the considered EoS, we evolve 5-6 configurations having mass in the range of $1.2 -2.0 M_{\odot}$. We compute $f$-mode frequency for each of the case and its dependence on the equation of state. We also compute tidal deformability for each of the case using perturbative approach. 

Our analysis show evidence of approximate universal relations between $f$-mode and other neutron star parameters such as tidal deformability, compactness etc. 
However, there is still a possibility to distinguish and constrain the possible stiff or soft equations of state based on the observed gravitational wave signal, as $f$-mode frequency value differs almost by $0.5$ kHz for soft EoS (APR4) and stiff EoS (DD2 or BHB) when the mass of the neutron star is $1.4 M_{\odot}$. The difference is higher for the more massive cases, in fact, for $M \geq 1.8 M_{\odot}$, frequency differs by $\approx 300$ Hz between DD2 (stiff EoS) and LS220 (having intermediate stiffness). Frequency computed are also well within the bounds reported by Ref.~\cite{hinderer_nature} through Bayesian estimates. 

Further, the computed frequencies match very well with the studies performed by other groups while studying the fundamental modes during the binary neutron star merger~\cite{steinhoff2021spin}, as well as the frequencies extracted through perturbative studies~\cite{Chirenti:2015PRD}.
These findings indicate that using single perturbed star simulations could give us similar results to that of a binary simulation, saving on computational time and resources for measuring $f$-mode frequency at inspiral phase. We also observe, similar tidal behaviour for SRO-APR and APR4 EoS, where the former is one of the most updated EoS in the family of APR EoS, hence it could be a good choice for binary merger simulations for future studies.

Our $f$-mode study of a perturbed neutron star in this paper is limited in many ways. 
Our numerical scheme for hydrodynamic evolution is limited to the second-order convergent finite volume method. In future studies, more advance numerical scheme such as discontinuous Galerkin method with higher accuracy and better efficiency can be used to study the perturbed neutron stars' evolution~\cite{Hebert-2018,Deppe-2021}.
We considered only non-rotating stars, and our EoS selection includes only one EoS with exotic matter i.e. BHB with hyperons.
In the follow-up studies, it is important to perform dynamical spacetime simulations of a perturbed rotating neutron star to investigate the relativistic corrections on $f$-mode frequency shift due to spin effects~\cite{steinhoff2021spin}.
Further, these studies can be enhanced by including a broad range of EoS with hyperons \cite{pradhan_chatterjee2021:PhysRevC.103.035810}, 
muons \cite{Wen-2019}, and also considering the parametrized EoS with piece-wise polytropic EoS \cite{Bauswein_2020_PhysRevLett.125.141103, miller2020:ApJ...888...12M} or with continuous sound speed  \cite{Boyle_nick_2020_PhysRevD.102.083027} to 
impose constraints on EoS, as are being considered in some of the recent studies.
In addition, we intend to consider longer evolution with higher resolution to have a better accuracy for the higher mode measurements, as well as for the $f$-mode damping time measurements. 
Similar asteroseismological studies can also be extended to theories of gravity beyond general relativity. Many models for compact stars in different types of modified gravity theories exist (for example, see Refs.~\cite{ref1, ref2, ref3, ref4}). Large-scale cosmological and gravitational wave observations may help probe these alternative gravity theories~\cite{Baker:2015}.


\section*{Acknowledgments}

The authors thank Sukanta Bose for frequent helpful discussions, scientific advice and comments over the entire course of this project.
We thank D. Radice for help and support with \texttt{WhiskyTHC} code. We are also grateful to Sukanta Bose and P. Pnigouras for their useful inputs on the manuscript.
S.S. acknowledges L. Baiotti, I. Hawke and International Centre for Theoretical Sciences (ICTS) for discussions during the program - Gravitational Wave Astrophysics (Code: ICTS/Prog-gws2020/05).
Part of this research is supported by the Navajbai Ratan Tata Trust and LIGO-India funds at IUCAA, India. F.H. acknowledges grant No. 2019/35/B/ST9/04000 from the Polish National Science Center, Poland.
S.S. also acknowledges support from the China Scholarship Council (CSC), Grant No. 2020GXZ016646.
The numerical simulations for this work were performed on Pegasus cluster, which is a part of the high performance computing (HPC) facility at The Inter-University Centre for Astronomy and Astrophysics (IUCAA), Pune, India.\\

\appendix

\section{Compactness - Tidal deformability universal relations}
\label{appendixA}

\begin{figure}
    \centering
    \includegraphics[width=0.55\textwidth]{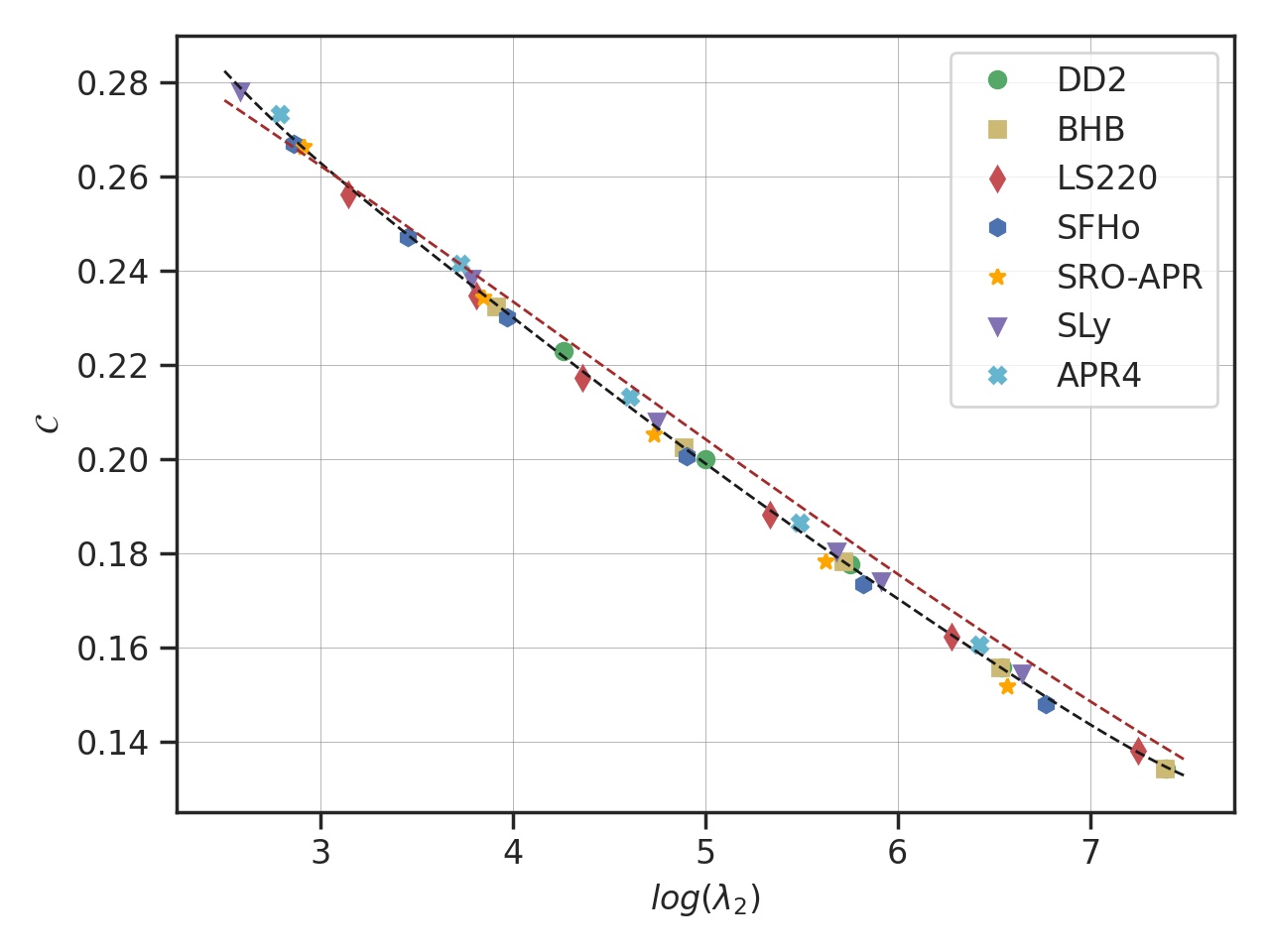}
    \caption{The universality observed between tidal deformability and compactness. Here, we compare our fit (black dashed line) with that of Ref.~\cite{Godzieba-etal:arxiv2012} (brown dashed line).}
    \label{fig:C-love_UR}
\end{figure}

For our compact star models, we also study another universal relation in addition to the discussion in Sec.~\ref{subsec:ur} between the compactness and tidal Love number using the initial data that we use in our simulations. We also use this as a check for our initial data against already established results \cite{Maselli2013:PhysRevD.88.023007, Godzieba-etal:arxiv2012}. The universal relations are given as,
\begin{equation}
\label{eqn:UR_c-l}
    \mathcal{C} = \sum_i b_i ~(log(\lambda_2))^i
\end{equation}
The universal relation between compactness and tidal deformability provides us the constraint on the radius of the star as a less compact star will be deformed more by a tidal potential for a given mass \cite{Godzieba-etal:arxiv2012}, providing a relation between radius and $\lambda_2$.
We present our obtained fit for $b_i$ in Tab.~\ref{tab:C-Love_fit}. In Fig.~\ref{fig:C-love_UR}, we compare our results to a recent study by \cite{Godzieba-etal:arxiv2012}. Our results are consistent with the results of Ref.~\cite{Maselli2013:PhysRevD.88.023007} as presented in Tab.~\ref{tab:C-Love_fit}.

\begin{table*}
    \centering
    \begin{tabular}{|c|c c|c c|}
    \hline
    Fit   & Maselli et al.~\cite{Maselli2013:PhysRevD.88.023007} & This work               & Godzieba et al.~\cite{Godzieba-etal:arxiv2012} & This work               \\ \hline
    $b_0$ & $3.71\times 10^{-1}$                       & $3.770 \times 10^{-1}$  & $3.388 \times 10^{-1}$              & $7.951 \times 10^{-1}$  \\ 
    $b_1$ & $-3.91 \times 10^{-2}$                     & $-4.137 \times 10^{-2}$ & $-2.30 \times 10^{-2}$              & $-5.839 \times 10^{-1}$ \\ 
    $b_2$ & $1.056 \times 10^{-3}$                     & $1.149 \times 10^{-3}$  & $-4.651 \times 10^{-4}$             & $2.877 \times 10^{-1}$  \\ 
    $b_3$ & -                                          & -                       & $-2.636 \times 10^{-4}$             & $-7.908 \times 10^{-2}$ \\ 
    $b_4$ & -                                          & -                       & $5.424 \times 10^{-5}$              & $1.205 \times 10^{-2}$  \\ 
    $b_5$ & -                                          & -                       & $-3.188 \times 10^{-6}$             & $-9.628 \times 10^{-4}$ \\ 
    $b_6$ & -                                          & -                       & $6.181 \times 10^{-8}$              & $3.155 \times 10^{-5}$  \\ \hline
    \end{tabular}
    \caption{Fitting parameters $b_i$ from Eq.~\ref{eqn:UR_c-l} for the $\mathcal{C}$-Love universal relations. Comparing this work with the coefficient values in Ref.~\cite{Maselli2013:PhysRevD.88.023007} and Ref.~\cite{Godzieba-etal:arxiv2012}. 
    The polytrope cases are excluded from $b_i$'s calculations. 
    }
    \label{tab:C-Love_fit}
\end{table*}

\bibliographystyle{elsarticle-num-names}
\bibliography{main}

\begin{thebibliography}{100}
\expandafter\ifx\csname natexlab\endcsname\relax\def\natexlab#1{#1}\fi
\providecommand{\url}[1]{\texttt{#1}}
\providecommand{\href}[2]{#2}
\providecommand{\path}[1]{#1}
\providecommand{\DOIprefix}{doi:}
\providecommand{\ArXivprefix}{arXiv:}
\providecommand{\URLprefix}{URL: }
\providecommand{\Pubmedprefix}{pmid:}
\providecommand{\doi}[1]{\href{http://dx.doi.org/#1}{\path{#1}}}
\providecommand{\Pubmed}[1]{\href{pmid:#1}{\path{#1}}}
\providecommand{\bibinfo}[2]{#2}
\ifx\xfnm\relax \def\xfnm[#1]{\unskip,\space#1}\fi
\bibitem[{Abbott et~al.(2017)}]{GW170817}
\bibinfo{author}{B.~P. Abbott}, et~al. (\bibinfo{collaboration}{LIGO
  Scientific, Virgo}),
\newblock \bibinfo{title}{{GW170817: Observation of Gravitational Waves from a
  Binary Neutron Star Inspiral}},
\newblock \bibinfo{journal}{Phys. Rev. Lett.} \bibinfo{volume}{119}
  (\bibinfo{year}{2017}) \bibinfo{pages}{161101}.
  \DOIprefix\doi{10.1103/PhysRevLett.119.161101}.
  \href{http://arxiv.org/abs/1710.05832}{{\tt arXiv:1710.05832}}.
\bibitem[{Abbott et~al.(2020)}]{GW190425}
\bibinfo{author}{B.~P. Abbott}, et~al. (\bibinfo{collaboration}{LIGO
  Scientific, Virgo}),
\newblock \bibinfo{title}{{GW190425: Observation of a Compact Binary
  Coalescence with Total Mass $\sim 3.4 M_{\odot}$}},
\newblock \bibinfo{journal}{Astrophys. J. Lett.} \bibinfo{volume}{892}
  (\bibinfo{year}{2020}) \bibinfo{pages}{L3}.
  \DOIprefix\doi{10.3847/2041-8213/ab75f5}.
  \href{http://arxiv.org/abs/2001.01761}{{\tt arXiv:2001.01761}}.
\bibitem[{Miller et~al.(2019)}]{miller2019:ApJ...887L..24M}
\bibinfo{author}{M.~C. Miller}, et~al.,
\newblock \bibinfo{title}{{PSR J0030+0451 Mass and Radius from $NICER$ Data and
  Implications for the Properties of Neutron Star Matter}},
\newblock \bibinfo{journal}{Astrophys. J. Lett.} \bibinfo{volume}{887}
  (\bibinfo{year}{2019}) \bibinfo{pages}{L24}.
  \DOIprefix\doi{10.3847/2041-8213/ab50c5}.
  \href{http://arxiv.org/abs/1912.05705}{{\tt arXiv:1912.05705}}.
\bibitem[{Bogdanov et~al.(2019{\natexlab{a}})}]{bogdanov2019a:ApJ...887L..25B}
\bibinfo{author}{S.~Bogdanov}, et~al.,
\newblock \bibinfo{title}{{Constraining the Neutron Star
  Mass\textendash{}Radius Relation and Dense Matter Equation of State with
  $NICER$. I. The Millisecond Pulsar X-Ray Data Set}},
\newblock \bibinfo{journal}{Astrophys. J. Lett.} \bibinfo{volume}{887}
  (\bibinfo{year}{2019}{\natexlab{a}}) \bibinfo{pages}{L25}.
  \DOIprefix\doi{10.3847/2041-8213/ab53eb}.
  \href{http://arxiv.org/abs/1912.05706}{{\tt arXiv:1912.05706}}.
\bibitem[{Bogdanov et~al.(2019{\natexlab{b}})}]{bogdanov2019b:ApJ...887L..26B}
\bibinfo{author}{S.~Bogdanov}, et~al.,
\newblock \bibinfo{title}{{Constraining the Neutron Star
  Mass\textendash{}Radius Relation and Dense Matter Equation of State with
  $NICER$. II. Emission from Hot Spots on a Rapidly Rotating Neutron Star}},
\newblock \bibinfo{journal}{Astrophys. J. Lett.} \bibinfo{volume}{887}
  (\bibinfo{year}{2019}{\natexlab{b}}) \bibinfo{pages}{L26}.
  \DOIprefix\doi{10.3847/2041-8213/ab5968}.
  \href{http://arxiv.org/abs/1912.05707}{{\tt arXiv:1912.05707}}.
\bibitem[{Raaijmakers et~al.(2020)}]{raaijmakers2020:ApJ...893L..21R}
\bibinfo{author}{G.~Raaijmakers}, et~al.,
\newblock \bibinfo{title}{{Constraining the dense matter equation of state with
  joint analysis of NICER and LIGO/Virgo measurements}},
\newblock \bibinfo{journal}{Astrophys. J. Lett.} \bibinfo{volume}{893}
  (\bibinfo{year}{2020}) \bibinfo{pages}{L21}.
  \DOIprefix\doi{10.3847/2041-8213/ab822f}.
  \href{http://arxiv.org/abs/1912.11031}{{\tt arXiv:1912.11031}}.
\bibitem[{Jiang et~al.(2020)Jiang, Tang, Wang, Fan, and
  Wei}]{Jiang2020:ApJ...892...55J}
\bibinfo{author}{J.-L. Jiang}, \bibinfo{author}{S.-P. Tang},
  \bibinfo{author}{Y.-Z. Wang}, \bibinfo{author}{Y.-Z. Fan},
  \bibinfo{author}{D.-M. Wei},
\newblock \bibinfo{title}{{PSR J0030+0451, GW170817 and the nuclear data: joint
  constraints on equation of state and bulk properties of neutron stars}},
\newblock \bibinfo{journal}{Astrophys. J.} \bibinfo{volume}{892}
  (\bibinfo{year}{2020}) \bibinfo{pages}{1}.
  \DOIprefix\doi{10.3847/1538-4357/ab77cf}.
  \href{http://arxiv.org/abs/1912.07467}{{\tt arXiv:1912.07467}}.
\bibitem[{Miller et~al.(2019)Miller, Chirenti, and
  Lamb}]{miller2020:ApJ...888...12M}
\bibinfo{author}{M.~C. Miller}, \bibinfo{author}{C.~Chirenti},
  \bibinfo{author}{F.~K. Lamb},
\newblock \bibinfo{title}{{Constraining the equation of state of high-density
  cold matter using nuclear and astronomical measurements}}
  (\bibinfo{year}{2019}). \DOIprefix\doi{10.3847/1538-4357/ab4ef9}.
  \href{http://arxiv.org/abs/1904.08907}{{\tt arXiv:1904.08907}}.
\bibitem[{Andersson and Kokkotas(1998)}]{andersson_kokkotas1998}
\bibinfo{author}{N.~Andersson}, \bibinfo{author}{K.~D. Kokkotas},
\newblock \bibinfo{title}{{Towards gravitational wave asteroseismology}},
\newblock \bibinfo{journal}{Mon. Not. Roy. Astron. Soc.} \bibinfo{volume}{299}
  (\bibinfo{year}{1998}) \bibinfo{pages}{1059--1068}.
  \DOIprefix\doi{10.1046/j.1365-8711.1998.01840.x}.
  \href{http://arxiv.org/abs/gr-qc/9711088}{{\tt arXiv:gr-qc/9711088}}.
\bibitem[{Allen et~al.(1998)Allen, Andersson, Kokkotas, and
  Schutz}]{gabrielle1998:PhysRevD.58.124012}
\bibinfo{author}{G.~Allen}, \bibinfo{author}{N.~Andersson},
  \bibinfo{author}{K.~D. Kokkotas}, \bibinfo{author}{B.~F. Schutz},
\newblock \bibinfo{title}{{Gravitational waves from pulsating stars: Evolving
  the perturbation equations for a relativistic star}},
\newblock \bibinfo{journal}{Phys. Rev. D} \bibinfo{volume}{58}
  (\bibinfo{year}{1998}) \bibinfo{pages}{124012}.
  \DOIprefix\doi{10.1103/PhysRevD.58.124012}.
  \href{http://arxiv.org/abs/gr-qc/9704023}{{\tt arXiv:gr-qc/9704023}}.
\bibitem[{Pons et~al.(2002)Pons, Berti, Gualtieri, Miniutti, and
  Ferrari}]{pons2002:PhysRevD.65.104021}
\bibinfo{author}{J.~A. Pons}, \bibinfo{author}{E.~Berti},
  \bibinfo{author}{L.~Gualtieri}, \bibinfo{author}{G.~Miniutti},
  \bibinfo{author}{V.~Ferrari},
\newblock \bibinfo{title}{{Gravitational signals emitted by a point mass
  orbiting a neutron star: Effects of stellar structure}},
\newblock \bibinfo{journal}{Phys. Rev. D} \bibinfo{volume}{65}
  (\bibinfo{year}{2002}) \bibinfo{pages}{104021}.
  \DOIprefix\doi{10.1103/PhysRevD.65.104021}.
  \href{http://arxiv.org/abs/gr-qc/0111104}{{\tt arXiv:gr-qc/0111104}}.
\bibitem[{Benhar et~al.(2004)Benhar, Ferrari, and
  Gualtieri}]{Benhar2004:PhysRevD.70.124015}
\bibinfo{author}{O.~Benhar}, \bibinfo{author}{V.~Ferrari},
  \bibinfo{author}{L.~Gualtieri},
\newblock \bibinfo{title}{{Gravitational wave asteroseismology revisited}},
\newblock \bibinfo{journal}{Phys. Rev. D} \bibinfo{volume}{70}
  (\bibinfo{year}{2004}) \bibinfo{pages}{124015}.
  \DOIprefix\doi{10.1103/PhysRevD.70.124015}.
  \href{http://arxiv.org/abs/astro-ph/0407529}{{\tt arXiv:astro-ph/0407529}}.
\bibitem[{Kokkotas(1995)}]{Kokkotas_1996}
\bibinfo{author}{K.~D. Kokkotas},
\newblock \bibinfo{title}{{Pulsating relativistic stars}},
\newblock in: \bibinfo{booktitle}{{Les Houches School of Physics: Astrophysical
  Sources of Gravitational Radiation}}, \bibinfo{year}{1995}, pp.
  \bibinfo{pages}{89--102}. \href{http://arxiv.org/abs/gr-qc/9603024}{{\tt
  arXiv:gr-qc/9603024}}.
\bibitem[{Lai(1994)}]{Lai1994}
\bibinfo{author}{D.~Lai},
\newblock \bibinfo{title}{{Resonant oscillations and tidal heating in
  coalescing binary neutron stars}},
\newblock \bibinfo{journal}{Mon. Not. Roy. Astron. Soc.} \bibinfo{volume}{270}
  (\bibinfo{year}{1994}) \bibinfo{pages}{611}.
  \DOIprefix\doi{10.1093/mnras/270.3.611}.
  \href{http://arxiv.org/abs/astro-ph/9404062}{{\tt arXiv:astro-ph/9404062}}.
\bibitem[{Lai and Wu(2006)}]{Lai2006}
\bibinfo{author}{D.~Lai}, \bibinfo{author}{Y.~Wu},
\newblock \bibinfo{title}{{Resonant Tidal Excitations of Inertial Modes in
  Coalescing Neutron Star Binaries}},
\newblock \bibinfo{journal}{Phys. Rev. D} \bibinfo{volume}{74}
  (\bibinfo{year}{2006}) \bibinfo{pages}{024007}.
  \DOIprefix\doi{10.1103/PhysRevD.74.024007}.
  \href{http://arxiv.org/abs/astro-ph/0604163}{{\tt arXiv:astro-ph/0604163}}.
\bibitem[{Xu and Lai(2017)}]{Xu-Lai2017}
\bibinfo{author}{W.~Xu}, \bibinfo{author}{D.~Lai},
\newblock \bibinfo{title}{{Resonant Tidal Excitation of Oscillation Modes in
  Merging Binary Neutron Stars: Inertial-Gravity Modes}},
\newblock \bibinfo{journal}{Phys. Rev. D} \bibinfo{volume}{96}
  (\bibinfo{year}{2017}) \bibinfo{pages}{083005}.
  \DOIprefix\doi{10.1103/PhysRevD.96.083005}.
  \href{http://arxiv.org/abs/1708.01839}{{\tt arXiv:1708.01839}}.
\bibitem[{Weinberg et~al.(2013)Weinberg, Arras, and Burkart}]{Weinberg_2013}
\bibinfo{author}{N.~N. Weinberg}, \bibinfo{author}{P.~Arras},
  \bibinfo{author}{J.~Burkart},
\newblock \bibinfo{title}{{An instability due to the nonlinear coupling of
  p-modes to g-modes: Implications for coalescing neutron star binaries}},
\newblock \bibinfo{journal}{Astrophys. J.} \bibinfo{volume}{769}
  (\bibinfo{year}{2013}) \bibinfo{pages}{121}.
  \DOIprefix\doi{10.1088/0004-637X/769/2/121}.
  \href{http://arxiv.org/abs/1302.2292}{{\tt arXiv:1302.2292}}.
\bibitem[{Weinberg(2016)}]{Weinberg_2016}
\bibinfo{author}{N.~N. Weinberg},
\newblock \bibinfo{title}{{Growth rate of the tidal p-mode g-mode instability
  in coalescing binary neutron stars}},
\newblock \bibinfo{journal}{Astrophys. J.} \bibinfo{volume}{819}
  (\bibinfo{year}{2016}) \bibinfo{pages}{109}.
  \DOIprefix\doi{10.3847/0004-637X/819/2/109}.
  \href{http://arxiv.org/abs/1509.06975}{{\tt arXiv:1509.06975}}.
\bibitem[{Zhou and Zhang(2017)}]{Zhou_2017}
\bibinfo{author}{Y.~Zhou}, \bibinfo{author}{F.~Zhang},
\newblock \bibinfo{title}{{Equation of State Dependence of Nonlinear Mode-tide
  Coupling in Coalescing Binary Neutron Stars}},
\newblock \bibinfo{journal}{Astrophys. J.} \bibinfo{volume}{849}
  (\bibinfo{year}{2017}) \bibinfo{pages}{114}.
  \DOIprefix\doi{10.3847/1538-4357/aa906e}.
  \href{http://arxiv.org/abs/1801.09675}{{\tt arXiv:1801.09675}}.
\bibitem[{Nouri et~al.(2021)Nouri, Bose, Duez, and Das}]{Nouri:2021}
\bibinfo{author}{F.~H. Nouri}, \bibinfo{author}{S.~Bose},
  \bibinfo{author}{M.~D. Duez}, \bibinfo{author}{A.~Das},
\newblock \bibinfo{title}{{Nonlinear mode-tide coupling in coalescing binary
  neutron stars with relativistic corrections}}  (\bibinfo{year}{2021}).
  \href{http://arxiv.org/abs/2107.13339}{{\tt arXiv:2107.13339}}.
\bibitem[{Andersson and Pnigouras(2021)}]{Andersson19:arxiv1905}
\bibinfo{author}{N.~Andersson}, \bibinfo{author}{P.~Pnigouras},
\newblock \bibinfo{title}{{The phenomenology of dynamical neutron star tides}},
\newblock \bibinfo{journal}{Mon. Not. Roy. Astron. Soc.} \bibinfo{volume}{503}
  (\bibinfo{year}{2021}) \bibinfo{pages}{533--539}.
  \DOIprefix\doi{10.1093/mnras/stab371}.
  \href{http://arxiv.org/abs/1905.00012}{{\tt arXiv:1905.00012}}.
\bibitem[{Andersson and Pnigouras(2020)}]{Andersson20:PhysRevD.101.083001}
\bibinfo{author}{N.~Andersson}, \bibinfo{author}{P.~Pnigouras},
\newblock \bibinfo{title}{{Exploring the effective tidal deformability of
  neutron stars}},
\newblock \bibinfo{journal}{Phys. Rev. D} \bibinfo{volume}{101}
  (\bibinfo{year}{2020}) \bibinfo{pages}{083001}.
  \DOIprefix\doi{10.1103/PhysRevD.101.083001}.
  \href{http://arxiv.org/abs/1906.08982}{{\tt arXiv:1906.08982}}.
\bibitem[{Pratten et~al.(2020)Pratten, Schmidt, and Hinderer}]{hinderer_nature}
\bibinfo{author}{G.~Pratten}, \bibinfo{author}{P.~Schmidt},
  \bibinfo{author}{T.~Hinderer},
\newblock \bibinfo{title}{{Gravitational-Wave Asteroseismology with Fundamental
  Modes from Compact Binary Inspirals}},
\newblock \bibinfo{journal}{Nature Commun.} \bibinfo{volume}{11}
  (\bibinfo{year}{2020}) \bibinfo{pages}{2553}.
  \DOIprefix\doi{10.1038/s41467-020-15984-5}.
  \href{http://arxiv.org/abs/1905.00817}{{\tt arXiv:1905.00817}}.
\bibitem[{Schmidt and Hinderer(2019)}]{Schmidt_Hinderer:2019_PhysRevD}
\bibinfo{author}{P.~Schmidt}, \bibinfo{author}{T.~Hinderer},
\newblock \bibinfo{title}{{Frequency domain model of $f$-mode dynamic tides in
  gravitational waveforms from compact binary inspirals}},
\newblock \bibinfo{journal}{Phys. Rev. D} \bibinfo{volume}{100}
  (\bibinfo{year}{2019}) \bibinfo{pages}{021501}.
  \DOIprefix\doi{10.1103/PhysRevD.100.021501}.
  \href{http://arxiv.org/abs/1905.00818}{{\tt arXiv:1905.00818}}.
\bibitem[{Ma et~al.(2020)Ma, Yu, and Chen}]{PhysRevD.101.123020}
\bibinfo{author}{S.~Ma}, \bibinfo{author}{H.~Yu}, \bibinfo{author}{Y.~Chen},
\newblock \bibinfo{title}{{Excitation of f-modes during mergers of spinning
  binary neutron star}},
\newblock \bibinfo{journal}{Phys. Rev. D} \bibinfo{volume}{101}
  (\bibinfo{year}{2020}) \bibinfo{pages}{123020}.
  \DOIprefix\doi{10.1103/PhysRevD.101.123020}.
  \href{http://arxiv.org/abs/2003.02373}{{\tt arXiv:2003.02373}}.
\bibitem[{Steinhoff et~al.(2021)Steinhoff, Hinderer, Dietrich, and
  Foucart}]{steinhoff2021spin}
\bibinfo{author}{J.~Steinhoff}, \bibinfo{author}{T.~Hinderer},
  \bibinfo{author}{T.~Dietrich}, \bibinfo{author}{F.~Foucart},
\newblock \bibinfo{title}{{Spin effects on neutron star fundamental-mode
  dynamical tides: Phenomenology and comparison to numerical simulations}},
\newblock \bibinfo{journal}{Phys. Rev. Res.} \bibinfo{volume}{3}
  (\bibinfo{year}{2021}) \bibinfo{pages}{033129}.
  \DOIprefix\doi{10.1103/PhysRevResearch.3.033129}.
  \href{http://arxiv.org/abs/2103.06100}{{\tt arXiv:2103.06100}}.
\bibitem[{Ng et~al.(2021)Ng, Cheong, Lin, and Li}]{Tjonnie-li:arxiv2012}
\bibinfo{author}{H.~H.-Y. Ng}, \bibinfo{author}{P.~C.-K. Cheong},
  \bibinfo{author}{L.-M. Lin}, \bibinfo{author}{T.~G.~F. Li},
\newblock \bibinfo{title}{{Gravitational-wave Asteroseismology with f-modes
  from Neutron Star Binaries at the Merger Phase}},
\newblock \bibinfo{journal}{Astrophys. J.} \bibinfo{volume}{915}
  (\bibinfo{year}{2021}) \bibinfo{pages}{108}.
  \DOIprefix\doi{10.3847/1538-4357/ac0141}.
  \href{http://arxiv.org/abs/2012.08263}{{\tt arXiv:2012.08263}}.
\bibitem[{Rezzolla and Takami(2016)}]{Rezzolla_2016:PhysRevD.93.124051}
\bibinfo{author}{L.~Rezzolla}, \bibinfo{author}{K.~Takami},
\newblock \bibinfo{title}{{Gravitational-wave signal from binary neutron stars:
  a systematic analysis of the spectral properties}},
\newblock \bibinfo{journal}{Phys. Rev. D} \bibinfo{volume}{93}
  (\bibinfo{year}{2016}) \bibinfo{pages}{124051}.
  \DOIprefix\doi{10.1103/PhysRevD.93.124051}.
  \href{http://arxiv.org/abs/1604.00246}{{\tt arXiv:1604.00246}}.
\bibitem[{Dietrich et~al.(2017{\natexlab{a}})Dietrich, Ujevic, Tichy, Bernuzzi,
  and Bruegmann}]{Dietrich_2017a:PhysRevD.95.024029}
\bibinfo{author}{T.~Dietrich}, \bibinfo{author}{M.~Ujevic},
  \bibinfo{author}{W.~Tichy}, \bibinfo{author}{S.~Bernuzzi},
  \bibinfo{author}{B.~Bruegmann},
\newblock \bibinfo{title}{{Gravitational waves and mass ejecta from binary
  neutron star mergers: Effect of the mass-ratio}},
\newblock \bibinfo{journal}{Phys. Rev. D} \bibinfo{volume}{95}
  (\bibinfo{year}{2017}{\natexlab{a}}) \bibinfo{pages}{024029}.
  \DOIprefix\doi{10.1103/PhysRevD.95.024029}.
  \href{http://arxiv.org/abs/1607.06636}{{\tt arXiv:1607.06636}}.
\bibitem[{Dietrich et~al.(2017{\natexlab{b}})Dietrich, Bernuzzi, Ujevic, and
  Tichy}]{Dietrich_2017b:PhysRevD.95.044045}
\bibinfo{author}{T.~Dietrich}, \bibinfo{author}{S.~Bernuzzi},
  \bibinfo{author}{M.~Ujevic}, \bibinfo{author}{W.~Tichy},
\newblock \bibinfo{title}{{Gravitational waves and mass ejecta from binary
  neutron star mergers: Effect of the stars' rotation}},
\newblock \bibinfo{journal}{Phys. Rev. D} \bibinfo{volume}{95}
  (\bibinfo{year}{2017}{\natexlab{b}}) \bibinfo{pages}{044045}.
  \DOIprefix\doi{10.1103/PhysRevD.95.044045}.
  \href{http://arxiv.org/abs/1611.07367}{{\tt arXiv:1611.07367}}.
\bibitem[{Lioutas et~al.(2021)Lioutas, Bauswein, and
  Stergioulas}]{Nick-etal:arxiv2102}
\bibinfo{author}{G.~Lioutas}, \bibinfo{author}{A.~Bauswein},
  \bibinfo{author}{N.~Stergioulas},
\newblock \bibinfo{title}{{Frequency deviations in universal relations of
  isolated neutron stars and postmerger remnants}},
\newblock \bibinfo{journal}{Phys. Rev. D} \bibinfo{volume}{104}
  (\bibinfo{year}{2021}) \bibinfo{pages}{043011}.
  \DOIprefix\doi{10.1103/PhysRevD.104.043011}.
  \href{http://arxiv.org/abs/2102.12455}{{\tt arXiv:2102.12455}}.
\bibitem[{Lioutas and Stergioulas(2018)}]{damping_nick_2018}
\bibinfo{author}{G.~Lioutas}, \bibinfo{author}{N.~Stergioulas},
\newblock \bibinfo{title}{{Universal and approximate relations for the
  gravitational-wave damping timescale of $f$-modes in neutron stars}},
\newblock \bibinfo{journal}{Gen. Rel. Grav.} \bibinfo{volume}{50}
  (\bibinfo{year}{2018}) \bibinfo{pages}{12}.
  \DOIprefix\doi{10.1007/s10714-017-2331-7}.
  \href{http://arxiv.org/abs/1709.10067}{{\tt arXiv:1709.10067}}.
\bibitem[{Maselli et~al.(2013)Maselli, Cardoso, Ferrari, Gualtieri, and
  Pani}]{Maselli2013:PhysRevD.88.023007}
\bibinfo{author}{A.~Maselli}, \bibinfo{author}{V.~Cardoso},
  \bibinfo{author}{V.~Ferrari}, \bibinfo{author}{L.~Gualtieri},
  \bibinfo{author}{P.~Pani},
\newblock \bibinfo{title}{{Equation-of-state-independent relations in neutron
  stars}},
\newblock \bibinfo{journal}{Phys. Rev. D} \bibinfo{volume}{88}
  (\bibinfo{year}{2013}) \bibinfo{pages}{023007}.
  \DOIprefix\doi{10.1103/PhysRevD.88.023007}.
  \href{http://arxiv.org/abs/1304.2052}{{\tt arXiv:1304.2052}}.
\bibitem[{Chan et~al.(2014)Chan, Sham, Leung, and
  Lin}]{ChanUR:PhysRevD.90.124023}
\bibinfo{author}{T.~K. Chan}, \bibinfo{author}{Y.~H. Sham},
  \bibinfo{author}{P.~T. Leung}, \bibinfo{author}{L.~M. Lin},
\newblock \bibinfo{title}{{Multipolar universal relations between f-mode
  frequency and tidal deformability of compact stars}},
\newblock \bibinfo{journal}{Phys. Rev. D} \bibinfo{volume}{90}
  (\bibinfo{year}{2014}) \bibinfo{pages}{124023}.
  \DOIprefix\doi{10.1103/PhysRevD.90.124023}.
  \href{http://arxiv.org/abs/1408.3789}{{\tt arXiv:1408.3789}}.
\bibitem[{Chirenti et~al.(2015)Chirenti, de~Souza, and
  Kastaun}]{Chirenti:2015PRD}
\bibinfo{author}{C.~Chirenti}, \bibinfo{author}{G.~H. de~Souza},
  \bibinfo{author}{W.~Kastaun},
\newblock \bibinfo{title}{{Fundamental oscillation modes of neutron stars:
  validity of universal relations}},
\newblock \bibinfo{journal}{Phys. Rev. D} \bibinfo{volume}{91}
  (\bibinfo{year}{2015}) \bibinfo{pages}{044034}.
  \DOIprefix\doi{10.1103/PhysRevD.91.044034}.
  \href{http://arxiv.org/abs/1501.02970}{{\tt arXiv:1501.02970}}.
\bibitem[{Yagi and Yunes(2017{\natexlab{a}})}]{Yagi:2016qmr}
\bibinfo{author}{K.~Yagi}, \bibinfo{author}{N.~Yunes},
\newblock \bibinfo{title}{{Approximate Universal Relations among Tidal
  Parameters for Neutron Star Binaries}},
\newblock \bibinfo{journal}{Class. Quant. Grav.} \bibinfo{volume}{34}
  (\bibinfo{year}{2017}{\natexlab{a}}) \bibinfo{pages}{015006}.
  \DOIprefix\doi{10.1088/1361-6382/34/1/015006}.
  \href{http://arxiv.org/abs/1608.06187}{{\tt arXiv:1608.06187}}.
\bibitem[{Yagi and Yunes(2017{\natexlab{b}})}]{Yagi:2016bkt}
\bibinfo{author}{K.~Yagi}, \bibinfo{author}{N.~Yunes},
\newblock \bibinfo{title}{{Approximate Universal Relations for Neutron Stars
  and Quark Stars}},
\newblock \bibinfo{journal}{Phys. Rept.} \bibinfo{volume}{681}
  (\bibinfo{year}{2017}{\natexlab{b}}) \bibinfo{pages}{1--72}.
  \DOIprefix\doi{10.1016/j.physrep.2017.03.002}.
  \href{http://arxiv.org/abs/1608.02582}{{\tt arXiv:1608.02582}}.
\bibitem[{Lindblom and Detweiler(1983)}]{Lindblom:1983ps}
\bibinfo{author}{L.~Lindblom}, \bibinfo{author}{S.~L. Detweiler},
\newblock \bibinfo{title}{{The quadrupole oscillations of neutron stars}},
\newblock \bibinfo{journal}{Astrophys. J. Suppl.} \bibinfo{volume}{53}
  (\bibinfo{year}{1983}) \bibinfo{pages}{73--92}.
  \DOIprefix\doi{10.1086/190884}.
\bibitem[{Detweiler and Lindblom(1985)}]{Detweiler_1985}
\bibinfo{author}{S.~L. Detweiler}, \bibinfo{author}{L.~Lindblom},
\newblock \bibinfo{title}{{On the nonradial pulsations of general relativistic
  stellar models}},
\newblock \bibinfo{journal}{Astrophys. J.} \bibinfo{volume}{292}
  (\bibinfo{year}{1985}) \bibinfo{pages}{12--15}.
  \DOIprefix\doi{10.1086/163127}.
\bibitem[{{Thorne} and {Campolattaro}(1967)}]{1967ApJ...149..591T}
\bibinfo{author}{K.~S. {Thorne}}, \bibinfo{author}{A.~{Campolattaro}},
\newblock \bibinfo{title}{{Non-Radial Pulsation of General-Relativistic Stellar
  Models. I. Analytic Analysis for L >= 2}},
\newblock \bibinfo{journal}{Astrophys. J.} \bibinfo{volume}{149}
  (\bibinfo{year}{1967}) \bibinfo{pages}{591}. \DOIprefix\doi{10.1086/149288}.
\bibitem[{Font et~al.(2000)Font, Stergioulas, and Kokkotas}]{Font_2000}
\bibinfo{author}{J.~A. Font}, \bibinfo{author}{N.~Stergioulas},
  \bibinfo{author}{K.~D. Kokkotas},
\newblock \bibinfo{title}{{Nonlinear hydrodynamical evolution of rotating
  relativistic stars: Numerical methods and code tests}},
\newblock \bibinfo{journal}{Mon. Not. Roy. Astron. Soc.} \bibinfo{volume}{313}
  (\bibinfo{year}{2000}) \bibinfo{pages}{678}.
  \DOIprefix\doi{10.1046/j.1365-8711.2000.03254.x}.
  \href{http://arxiv.org/abs/gr-qc/9908010}{{\tt arXiv:gr-qc/9908010}}.
\bibitem[{Font et~al.(2001)Font, Dimmelmeier, Gupta, and
  Stergioulas}]{font_rotating}
\bibinfo{author}{J.~A. Font}, \bibinfo{author}{H.~Dimmelmeier},
  \bibinfo{author}{A.~Gupta}, \bibinfo{author}{N.~Stergioulas},
\newblock \bibinfo{title}{{Axisymmetric modes of rotating relativistic stars in
  the Cowling approximation}},
\newblock \bibinfo{journal}{Mon. Not. Roy. Astron. Soc.} \bibinfo{volume}{325}
  (\bibinfo{year}{2001}) \bibinfo{pages}{1463}.
  \DOIprefix\doi{10.1046/j.1365-8711.2001.04555.x}.
  \href{http://arxiv.org/abs/astro-ph/0012477}{{\tt arXiv:astro-ph/0012477}}.
\bibitem[{Shibata and Karino(2004)}]{Shibata2004}
\bibinfo{author}{M.~Shibata}, \bibinfo{author}{S.~Karino},
\newblock \bibinfo{title}{{Numerical evolution of secular bar-mode instability
  induced by the gravitational radiation reaction in rapidly rotating neutron
  stars}},
\newblock \bibinfo{journal}{Phys. Rev. D} \bibinfo{volume}{70}
  (\bibinfo{year}{2004}) \bibinfo{pages}{084022}.
  \DOIprefix\doi{10.1103/PhysRevD.70.084022}.
  \href{http://arxiv.org/abs/astro-ph/0408016}{{\tt arXiv:astro-ph/0408016}}.
\bibitem[{Kastaun et~al.(2010)Kastaun, Willburger, and
  Kokkotas}]{Kastaun_2010_CA:PhysRevD.82.104036}
\bibinfo{author}{W.~Kastaun}, \bibinfo{author}{B.~Willburger},
  \bibinfo{author}{K.~D. Kokkotas},
\newblock \bibinfo{title}{{On the saturation amplitude of the f-mode
  instability}},
\newblock \bibinfo{journal}{Phys. Rev. D} \bibinfo{volume}{82}
  (\bibinfo{year}{2010}) \bibinfo{pages}{104036}.
  \DOIprefix\doi{10.1103/PhysRevD.82.104036}.
  \href{http://arxiv.org/abs/1006.3885}{{\tt arXiv:1006.3885}}.
\bibitem[{{H{\'e}bert} et~al.(2018){H{\'e}bert}, {Kidder}, and
  {Teukolsky}}]{Hebert-2018}
\bibinfo{author}{F.~{H{\'e}bert}}, \bibinfo{author}{L.~E. {Kidder}},
  \bibinfo{author}{S.~A. {Teukolsky}},
\newblock \bibinfo{title}{{General-relativistic neutron star evolutions with
  the discontinuous Galerkin method}},
\newblock \bibinfo{journal}{Phys. Rev. D} \bibinfo{volume}{98}
  (\bibinfo{year}{2018}) \bibinfo{pages}{044041}.
  \DOIprefix\doi{10.1103/PhysRevD.98.044041}.
  \href{http://arxiv.org/abs/1804.02003}{{\tt arXiv:1804.02003}}.
\bibitem[{De~Pietri et~al.(2014)De~Pietri, Feo, Franci, and
  L\"offler}]{DePietri2014}
\bibinfo{author}{R.~De~Pietri}, \bibinfo{author}{A.~Feo},
  \bibinfo{author}{L.~Franci}, \bibinfo{author}{F.~L\"offler},
\newblock \bibinfo{title}{{Neutron Star instabilities in full General
  Relativity using a $\Gamma=2.75$ ideal fluid}},
\newblock \bibinfo{journal}{Phys. Rev. D} \bibinfo{volume}{90}
  (\bibinfo{year}{2014}) \bibinfo{pages}{024034}.
  \DOIprefix\doi{10.1103/PhysRevD.90.024034}.
  \href{http://arxiv.org/abs/1403.8066}{{\tt arXiv:1403.8066}}.
\bibitem[{Dimmelmeier et~al.(2006)Dimmelmeier, Stergioulas, and
  Font}]{Dimmelmeier2006:MNRAS}
\bibinfo{author}{H.~Dimmelmeier}, \bibinfo{author}{N.~Stergioulas},
  \bibinfo{author}{J.~A. Font},
\newblock \bibinfo{title}{{Non-linear axisymmetric pulsations of rotating
  relativistic stars in the conformal flatness approximation}},
\newblock \bibinfo{journal}{Mon. Not. Roy. Astron. Soc.} \bibinfo{volume}{368}
  (\bibinfo{year}{2006}) \bibinfo{pages}{1609--1630}.
  \DOIprefix\doi{10.1111/j.1365-2966.2006.10274.x}.
  \href{http://arxiv.org/abs/astro-ph/0511394}{{\tt arXiv:astro-ph/0511394}}.
\bibitem[{Bucciantini and Del~Zanna(2011)}]{Bucciantini_2011:aap}
\bibinfo{author}{N.~Bucciantini}, \bibinfo{author}{L.~Del~Zanna},
\newblock \bibinfo{title}{{GRMHD in axisymmetric dynamical spacetimes: the
  X-ECHO code}},
\newblock \bibinfo{journal}{Astron. Astrophys.} \bibinfo{volume}{528}
  (\bibinfo{year}{2011}) \bibinfo{pages}{A101}.
  \DOIprefix\doi{10.1051/0004-6361/201015945}.
  \href{http://arxiv.org/abs/1010.3532}{{\tt arXiv:1010.3532}}.
\bibitem[{Pili et~al.(2014)Pili, Bucciantini, and Del~Zanna}]{Pili_2014:mnras}
\bibinfo{author}{A.~G. Pili}, \bibinfo{author}{N.~Bucciantini},
  \bibinfo{author}{L.~Del~Zanna},
\newblock \bibinfo{title}{{Axisymmetric equilibrium models for magnetized
  neutron stars in General Relativity under the Conformally Flat Condition}},
\newblock \bibinfo{journal}{Mon. Not. Roy. Astron. Soc.} \bibinfo{volume}{439}
  (\bibinfo{year}{2014}) \bibinfo{pages}{3541--3563}.
  \DOIprefix\doi{10.1093/mnras/stu215}.
  \href{http://arxiv.org/abs/1401.4308}{{\tt arXiv:1401.4308}}.
\bibitem[{Rosofsky et~al.(2019)Rosofsky, Gold, Chirenti, Huerta, and
  Miller}]{rosofsky}
\bibinfo{author}{S.~Rosofsky}, \bibinfo{author}{R.~Gold},
  \bibinfo{author}{C.~Chirenti}, \bibinfo{author}{E.~A. Huerta},
  \bibinfo{author}{M.~C. Miller},
\newblock \bibinfo{title}{{Probing neutron star structure via f-mode
  oscillations and damping in dynamical spacetime models}},
\newblock \bibinfo{journal}{Phys. Rev. D} \bibinfo{volume}{99}
  (\bibinfo{year}{2019}) \bibinfo{pages}{084024}.
  \DOIprefix\doi{10.1103/PhysRevD.99.084024}.
  \href{http://arxiv.org/abs/1812.06126}{{\tt arXiv:1812.06126}}.
\bibitem[{Bauswein et~al.(2020)Bauswein, Blacker, Vijayan, Stergioulas,
  Chatziioannou, Clark, Bastian, Blaschke, Cierniak, and
  Fischer}]{Bauswein_2020_PhysRevLett.125.141103}
\bibinfo{author}{A.~Bauswein}, \bibinfo{author}{S.~Blacker},
  \bibinfo{author}{V.~Vijayan}, \bibinfo{author}{N.~Stergioulas},
  \bibinfo{author}{K.~Chatziioannou}, \bibinfo{author}{J.~A. Clark},
  \bibinfo{author}{N.-U.~F. Bastian}, \bibinfo{author}{D.~B. Blaschke},
  \bibinfo{author}{M.~Cierniak}, \bibinfo{author}{T.~Fischer},
\newblock \bibinfo{title}{{Equation of state constraints from the threshold
  binary mass for prompt collapse of neutron star mergers}},
\newblock \bibinfo{journal}{Phys. Rev. Lett.} \bibinfo{volume}{125}
  (\bibinfo{year}{2020}) \bibinfo{pages}{141103}.
  \DOIprefix\doi{10.1103/PhysRevLett.125.141103}.
  \href{http://arxiv.org/abs/2004.00846}{{\tt arXiv:2004.00846}}.
\bibitem[{O'Boyle et~al.(2020)O'Boyle, Markakis, Stergioulas, and
  Read}]{Boyle_nick_2020_PhysRevD.102.083027}
\bibinfo{author}{M.~F. O'Boyle}, \bibinfo{author}{C.~Markakis},
  \bibinfo{author}{N.~Stergioulas}, \bibinfo{author}{J.~S. Read},
\newblock \bibinfo{title}{{Parametrized equation of state for neutron star
  matter with continuous sound speed}},
\newblock \bibinfo{journal}{Phys. Rev. D} \bibinfo{volume}{102}
  (\bibinfo{year}{2020}) \bibinfo{pages}{083027}.
  \DOIprefix\doi{10.1103/PhysRevD.102.083027}.
  \href{http://arxiv.org/abs/2008.03342}{{\tt arXiv:2008.03342}}.
\bibitem[{Alcubierre(2008)}]{Alcubierre_2008}
\bibinfo{author}{M.~Alcubierre}, \bibinfo{title}{Introduction to 3+1 Numerical
  Relativity}, \bibinfo{publisher}{Oxford University Press},
  \bibinfo{address}{Oxford}, \bibinfo{year}{2008}.
\bibitem[{Rezzolla and Zanotti(2013)}]{Rezzolla_Zanotti_2013}
\bibinfo{author}{L.~Rezzolla}, \bibinfo{author}{O.~Zanotti},
  \bibinfo{title}{Relativistic Hydrodynamics}, \bibinfo{publisher}{Oxford
  University Press}, \bibinfo{address}{Oxford}, \bibinfo{year}{2013}.
\bibitem[{Baumgarte and Shapiro(2010)}]{Baumgarte_Shapiro_2010}
\bibinfo{author}{T.~Baumgarte}, \bibinfo{author}{S.~Shapiro},
  \bibinfo{title}{Numerical Relativity: Solving Einstein’s Equations on the
  Computer}, \bibinfo{publisher}{Cambridge University Press},
  \bibinfo{year}{2010}.
\bibitem[{Shibata(2015)}]{Shibata_2015}
\bibinfo{author}{M.~Shibata}, \bibinfo{title}{Numerical Relativity},
  \bibinfo{publisher}{World Scientific Publishing Company},
  \bibinfo{year}{2015}.
\bibitem[{Nakamura et~al.(1987)Nakamura, Oohara, and Kojima}]{oohara_kojima}
\bibinfo{author}{T.~Nakamura}, \bibinfo{author}{K.~Oohara},
  \bibinfo{author}{Y.~Kojima},
\newblock \bibinfo{title}{{General Relativistic Collapse to Black Holes and
  Gravitational Waves from Black Holes}},
\newblock \bibinfo{journal}{Prog. Theor. Phys. Suppl.} \bibinfo{volume}{90}
  (\bibinfo{year}{1987}) \bibinfo{pages}{1--218}.
  \DOIprefix\doi{10.1143/PTPS.90.1}.
\bibitem[{Shibata and Nakamura(1995)}]{shibata_nakamura}
\bibinfo{author}{M.~Shibata}, \bibinfo{author}{T.~Nakamura},
\newblock \bibinfo{title}{{Evolution of three-dimensional gravitational waves:
  Harmonic slicing case}},
\newblock \bibinfo{journal}{Phys. Rev. D} \bibinfo{volume}{52}
  (\bibinfo{year}{1995}) \bibinfo{pages}{5428--5444}.
  \DOIprefix\doi{10.1103/PhysRevD.52.5428}.
\bibitem[{Baumgarte and Shapiro(1998)}]{baumgarte_shapiro}
\bibinfo{author}{T.~W. Baumgarte}, \bibinfo{author}{S.~L. Shapiro},
\newblock \bibinfo{title}{{On the numerical integration of Einstein's field
  equations}},
\newblock \bibinfo{journal}{Phys. Rev. D} \bibinfo{volume}{59}
  (\bibinfo{year}{1998}) \bibinfo{pages}{024007}.
  \DOIprefix\doi{10.1103/PhysRevD.59.024007}.
  \href{http://arxiv.org/abs/gr-qc/9810065}{{\tt arXiv:gr-qc/9810065}}.
\bibitem[{Alcubierre et~al.(2000)Alcubierre, Allen, Bruegmann, Dramlitsch,
  Font, Papadopoulos, Seidel, Stergioulas, Suen, and
  Takahashi}]{alcubierre_bruegmann}
\bibinfo{author}{M.~Alcubierre}, \bibinfo{author}{G.~Allen},
  \bibinfo{author}{B.~Bruegmann}, \bibinfo{author}{T.~Dramlitsch},
  \bibinfo{author}{J.~A. Font}, \bibinfo{author}{P.~Papadopoulos},
  \bibinfo{author}{E.~Seidel}, \bibinfo{author}{N.~Stergioulas},
  \bibinfo{author}{W.-M. Suen}, \bibinfo{author}{R.~Takahashi},
\newblock \bibinfo{title}{{Towards a stable numerical evolution of strongly
  gravitating systems in general relativity: The Conformal treatments}},
\newblock \bibinfo{journal}{Phys. Rev. D} \bibinfo{volume}{62}
  (\bibinfo{year}{2000}) \bibinfo{pages}{044034}.
  \DOIprefix\doi{10.1103/PhysRevD.62.044034}.
  \href{http://arxiv.org/abs/gr-qc/0003071}{{\tt arXiv:gr-qc/0003071}}.
\bibitem[{Arnowitt et~al.(1959)Arnowitt, Deser, and Misner}]{adm1}
\bibinfo{author}{R.~L. Arnowitt}, \bibinfo{author}{S.~Deser},
  \bibinfo{author}{C.~W. Misner},
\newblock \bibinfo{title}{{Dynamical Structure and Definition of Energy in
  General Relativity}},
\newblock \bibinfo{journal}{Phys. Rev.} \bibinfo{volume}{116}
  (\bibinfo{year}{1959}) \bibinfo{pages}{1322--1330}.
  \DOIprefix\doi{10.1103/PhysRev.116.1322}.
\bibitem[{Arnowitt et~al.(2008)Arnowitt, Deser, and Misner}]{adm2}
\bibinfo{author}{R.~L. Arnowitt}, \bibinfo{author}{S.~Deser},
  \bibinfo{author}{C.~W. Misner},
\newblock \bibinfo{title}{{The Dynamics of general relativity}},
\newblock \bibinfo{journal}{Gen. Rel. Grav.} \bibinfo{volume}{40}
  (\bibinfo{year}{2008}) \bibinfo{pages}{1997--2027}.
  \DOIprefix\doi{10.1007/s10714-008-0661-1}.
  \href{http://arxiv.org/abs/gr-qc/0405109}{{\tt arXiv:gr-qc/0405109}}.
\bibitem[{Baiotti and Rezzolla(2017)}]{Baiotti_Rezzolla_2017_review}
\bibinfo{author}{L.~Baiotti}, \bibinfo{author}{L.~Rezzolla},
\newblock \bibinfo{title}{{Binary neutron star mergers: a review of
  Einstein\textquoteright{}s richest laboratory}},
\newblock \bibinfo{journal}{Rept. Prog. Phys.} \bibinfo{volume}{80}
  (\bibinfo{year}{2017}) \bibinfo{pages}{096901}.
  \DOIprefix\doi{10.1088/1361-6633/aa67bb}.
  \href{http://arxiv.org/abs/1607.03540}{{\tt arXiv:1607.03540}}.
\bibitem[{Brown et~al.(2009)Brown, Diener, Sarbach, Schnetter, and
  Tiglio}]{ml_bssn}
\bibinfo{author}{J.~D. Brown}, \bibinfo{author}{P.~Diener},
  \bibinfo{author}{O.~Sarbach}, \bibinfo{author}{E.~Schnetter},
  \bibinfo{author}{M.~Tiglio},
\newblock \bibinfo{title}{{Turduckening black holes: An Analytical and
  computational study}},
\newblock \bibinfo{journal}{Phys. Rev. D} \bibinfo{volume}{79}
  (\bibinfo{year}{2009}) \bibinfo{pages}{044023}.
  \DOIprefix\doi{10.1103/PhysRevD.79.044023}.
  \href{http://arxiv.org/abs/0809.3533}{{\tt arXiv:0809.3533}}.
\bibitem[{Babiuc-Hamilton et~al.(2019)}]{etk2019}
\bibinfo{author}{M.~Babiuc-Hamilton}, et~al., \bibinfo{title}{The einstein
  toolkit}, \bibinfo{year}{2019}. \DOIprefix\doi{10.5281/zenodo.3522086},
  \bibinfo{note}{to find out more, visit http://einsteintoolkit.org}.
\bibitem[{Goodale et~al.(2003)Goodale, Allen, Lanfermann, Mass{\'o}, Radke,
  Seidel, and Shalf}]{cactus}
\bibinfo{author}{T.~Goodale}, \bibinfo{author}{G.~Allen},
  \bibinfo{author}{G.~Lanfermann}, \bibinfo{author}{J.~Mass{\'o}},
  \bibinfo{author}{T.~Radke}, \bibinfo{author}{E.~Seidel},
  \bibinfo{author}{J.~Shalf},
\newblock \bibinfo{title}{The {Cactus} framework and toolkit: Design and
  applications},
\newblock in: \bibinfo{booktitle}{Vector and Parallel Processing --
  VECPAR'2002, 5th International Conference, Lecture Notes in Computer
  Science}, \bibinfo{publisher}{Springer}, \bibinfo{address}{Berlin},
  \bibinfo{year}{2003}. \URLprefix \url{http://edoc.mpg.de/3341}.
\bibitem[{Schnetter et~al.(2006)Schnetter, Diener, Dorband, and
  Tiglio}]{carpet1}
\bibinfo{author}{E.~Schnetter}, \bibinfo{author}{P.~Diener},
  \bibinfo{author}{E.~N. Dorband}, \bibinfo{author}{M.~Tiglio},
\newblock \bibinfo{title}{{A Multi-block infrastructure for three-dimensional
  time-dependent numerical relativity}},
\newblock \bibinfo{journal}{Class. Quant. Grav.} \bibinfo{volume}{23}
  (\bibinfo{year}{2006}) \bibinfo{pages}{S553--S578}.
  \DOIprefix\doi{10.1088/0264-9381/23/16/S14}.
  \href{http://arxiv.org/abs/gr-qc/0602104}{{\tt arXiv:gr-qc/0602104}}.
\bibitem[{Schnetter et~al.(2004)Schnetter, Hawley, and Hawke}]{carpet2}
\bibinfo{author}{E.~Schnetter}, \bibinfo{author}{S.~H. Hawley},
  \bibinfo{author}{I.~Hawke},
\newblock \bibinfo{title}{{Evolutions in 3-D numerical relativity using fixed
  mesh refinement}},
\newblock \bibinfo{journal}{Class. Quant. Grav.} \bibinfo{volume}{21}
  (\bibinfo{year}{2004}) \bibinfo{pages}{1465--1488}.
  \DOIprefix\doi{10.1088/0264-9381/21/6/014}.
  \href{http://arxiv.org/abs/gr-qc/0310042}{{\tt arXiv:gr-qc/0310042}}.
\bibitem[{Font(2008)}]{Font_living_review}
\bibinfo{author}{J.~A. Font},
\newblock \bibinfo{title}{{Numerical Hydrodynamics and Magnetohydrodynamics in
  General Relativity}},
\newblock \bibinfo{journal}{Living Rev. Rel.} \bibinfo{volume}{11}
  (\bibinfo{year}{2008}) \bibinfo{pages}{7}.
  \DOIprefix\doi{10.12942/lrr-2008-7}.
\bibitem[{Radice et~al.(2014{\natexlab{a}})Radice, Rezzolla, and
  Galeazzi}]{thc1}
\bibinfo{author}{D.~Radice}, \bibinfo{author}{L.~Rezzolla},
  \bibinfo{author}{F.~Galeazzi},
\newblock \bibinfo{title}{{Beyond second-order convergence in simulations of
  binary neutron stars in full general-relativity}},
\newblock \bibinfo{journal}{Mon. Not. Roy. Astron. Soc.} \bibinfo{volume}{437}
  (\bibinfo{year}{2014}{\natexlab{a}}) \bibinfo{pages}{L46--L50}.
  \DOIprefix\doi{10.1093/mnrasl/slt137}.
  \href{http://arxiv.org/abs/1306.6052}{{\tt arXiv:1306.6052}}.
\bibitem[{Radice et~al.(2014{\natexlab{b}})Radice, Rezzolla, and
  Galeazzi}]{thc2}
\bibinfo{author}{D.~Radice}, \bibinfo{author}{L.~Rezzolla},
  \bibinfo{author}{F.~Galeazzi},
\newblock \bibinfo{title}{{High-Order Fully General-Relativistic Hydrodynamics:
  new Approaches and Tests}},
\newblock \bibinfo{journal}{Class. Quant. Grav.} \bibinfo{volume}{31}
  (\bibinfo{year}{2014}{\natexlab{b}}) \bibinfo{pages}{075012}.
  \DOIprefix\doi{10.1088/0264-9381/31/7/075012}.
  \href{http://arxiv.org/abs/1312.5004}{{\tt arXiv:1312.5004}}.
\bibitem[{Radice and Rezzolla(2012)}]{thc3}
\bibinfo{author}{D.~Radice}, \bibinfo{author}{L.~Rezzolla},
\newblock \bibinfo{title}{{THC: a new high-order finite-difference
  high-resolution shock-capturing code for special-relativistic
  hydrodynamics}},
\newblock \bibinfo{journal}{Astron. Astrophys.} \bibinfo{volume}{547}
  (\bibinfo{year}{2012}) \bibinfo{pages}{A26}.
  \DOIprefix\doi{10.1051/0004-6361/201219735}.
  \href{http://arxiv.org/abs/1206.6502}{{\tt arXiv:1206.6502}}.
\bibitem[{Lattimer and Swesty(1991)}]{Lattimer:1991}
\bibinfo{author}{J.~M. Lattimer}, \bibinfo{author}{F.~D. Swesty},
\newblock \bibinfo{title}{{A Generalized equation of state for hot, dense
  matter}},
\newblock \bibinfo{journal}{Nucl. Phys. A} \bibinfo{volume}{535}
  (\bibinfo{year}{1991}) \bibinfo{pages}{331--376}.
  \DOIprefix\doi{10.1016/0375-9474(91)90452-C}.
\bibitem[{Hempel et~al.(2012)Hempel, Fischer, Schaffner-Bielich, and
  Liebendorfer}]{Hempel_2012}
\bibinfo{author}{M.~Hempel}, \bibinfo{author}{T.~Fischer},
  \bibinfo{author}{J.~Schaffner-Bielich}, \bibinfo{author}{M.~Liebendorfer},
\newblock \bibinfo{title}{{New Equations of State in Simulations of
  Core-Collapse Supernovae}},
\newblock \bibinfo{journal}{Astrophys. J.} \bibinfo{volume}{748}
  (\bibinfo{year}{2012}) \bibinfo{pages}{70}.
  \DOIprefix\doi{10.1088/0004-637X/748/1/70}.
  \href{http://arxiv.org/abs/1108.0848}{{\tt arXiv:1108.0848}}.
\bibitem[{Steiner et~al.(2013)Steiner, Hempel, and Fischer}]{Steiner_2013}
\bibinfo{author}{A.~W. Steiner}, \bibinfo{author}{M.~Hempel},
  \bibinfo{author}{T.~Fischer},
\newblock \bibinfo{title}{{Core-collapse supernova equations of state based on
  neutron star observations}},
\newblock \bibinfo{journal}{Astrophys. J.} \bibinfo{volume}{774}
  (\bibinfo{year}{2013}) \bibinfo{pages}{17}.
  \DOIprefix\doi{10.1088/0004-637X/774/1/17}.
  \href{http://arxiv.org/abs/1207.2184}{{\tt arXiv:1207.2184}}.
\bibitem[{Banik et~al.(2014)Banik, Hempel, and Bandyopadhyay}]{Banik_2014}
\bibinfo{author}{S.~Banik}, \bibinfo{author}{M.~Hempel},
  \bibinfo{author}{D.~Bandyopadhyay},
\newblock \bibinfo{title}{{New Hyperon Equations of State for Supernovae and
  Neutron Stars in Density-dependent Hadron Field Theory}},
\newblock \bibinfo{journal}{Astrophys. J. Suppl.} \bibinfo{volume}{214}
  (\bibinfo{year}{2014}) \bibinfo{pages}{22}.
  \DOIprefix\doi{10.1088/0067-0049/214/2/22}.
  \href{http://arxiv.org/abs/1404.6173}{{\tt arXiv:1404.6173}}.
\bibitem[{Chabanat et~al.(1998)Chabanat, Bonche, Haensel, Meyer, and
  Schaeffer}]{Chabanat:1997un}
\bibinfo{author}{E.~Chabanat}, \bibinfo{author}{P.~Bonche},
  \bibinfo{author}{P.~Haensel}, \bibinfo{author}{J.~Meyer},
  \bibinfo{author}{R.~Schaeffer},
\newblock \bibinfo{title}{{A Skyrme parametrization from subnuclear to neutron
  star densities. 2. Nuclei far from stablities}},
\newblock \bibinfo{journal}{Nucl. Phys. A} \bibinfo{volume}{635}
  (\bibinfo{year}{1998}) \bibinfo{pages}{231--256}.
  \DOIprefix\doi{10.1016/S0375-9474(98)00180-8}, \bibinfo{note}{[Erratum:
  Nucl.Phys.A 643, 441--441 (1998)]}.
\bibitem[{Akmal et~al.(1998)Akmal, Pandharipande, and Ravenhall}]{akmal1998}
\bibinfo{author}{A.~Akmal}, \bibinfo{author}{V.~R. Pandharipande},
  \bibinfo{author}{D.~G. Ravenhall},
\newblock \bibinfo{title}{{The Equation of state of nucleon matter and neutron
  star structure}},
\newblock \bibinfo{journal}{Phys. Rev. C} \bibinfo{volume}{58}
  (\bibinfo{year}{1998}) \bibinfo{pages}{1804--1828}.
  \DOIprefix\doi{10.1103/PhysRevC.58.1804}.
  \href{http://arxiv.org/abs/nucl-th/9804027}{{\tt arXiv:nucl-th/9804027}}.
\bibitem[{Schneider et~al.(2019)Schneider, Constantinou, Muccioli, and
  Prakash}]{Schneider2019}
\bibinfo{author}{A.~S. Schneider}, \bibinfo{author}{C.~Constantinou},
  \bibinfo{author}{B.~Muccioli}, \bibinfo{author}{M.~Prakash},
\newblock \bibinfo{title}{{Akmal-Pandharipande-Ravenhall equation of state for
  simulations of supernovae, neutron stars, and binary mergers}},
\newblock \bibinfo{journal}{Phys. Rev. C} \bibinfo{volume}{100}
  (\bibinfo{year}{2019}) \bibinfo{pages}{025803}.
  \DOIprefix\doi{10.1103/PhysRevC.100.025803}.
  \href{http://arxiv.org/abs/1901.09652}{{\tt arXiv:1901.09652}}.
\bibitem[{Takami et~al.(2015)Takami, Rezzolla, and Baiotti}]{takami_hybrid}
\bibinfo{author}{K.~Takami}, \bibinfo{author}{L.~Rezzolla},
  \bibinfo{author}{L.~Baiotti},
\newblock \bibinfo{title}{{Spectral properties of the post-merger
  gravitational-wave signal from binary neutron stars}},
\newblock \bibinfo{journal}{Phys. Rev. D} \bibinfo{volume}{91}
  (\bibinfo{year}{2015}) \bibinfo{pages}{064001}.
  \DOIprefix\doi{10.1103/PhysRevD.91.064001}.
  \href{http://arxiv.org/abs/1412.3240}{{\tt arXiv:1412.3240}}.
\bibitem[{Figura et~al.(2020)Figura, Lu, Burgio, Li, and
  Schulze}]{figura2020hybrid}
\bibinfo{author}{A.~Figura}, \bibinfo{author}{J.~J. Lu}, \bibinfo{author}{G.~F.
  Burgio}, \bibinfo{author}{Z.~H. Li}, \bibinfo{author}{H.~J. Schulze},
\newblock \bibinfo{title}{{Hybrid equation of state approach in binary
  neutron-star merger simulations}},
\newblock \bibinfo{journal}{Phys. Rev. D} \bibinfo{volume}{102}
  (\bibinfo{year}{2020}) \bibinfo{pages}{043006}.
  \DOIprefix\doi{10.1103/PhysRevD.102.043006}.
  \href{http://arxiv.org/abs/2005.08691}{{\tt arXiv:2005.08691}}.
\bibitem[{Cromartie et~al.(2019)}]{Cromartie2020}
\bibinfo{author}{H.~T. Cromartie}, et~al. (\bibinfo{collaboration}{NANOGrav}),
\newblock \bibinfo{title}{{Relativistic Shapiro delay measurements of an
  extremely massive millisecond pulsar}},
\newblock \bibinfo{journal}{Nature Astron.} \bibinfo{volume}{4}
  (\bibinfo{year}{2019}) \bibinfo{pages}{72--76}.
  \DOIprefix\doi{10.1038/s41550-019-0880-2}.
  \href{http://arxiv.org/abs/1904.06759}{{\tt arXiv:1904.06759}}.
\bibitem[{Abbott et~al.(2018)}]{EoS_GW170817}
\bibinfo{author}{B.~P. Abbott}, et~al. (\bibinfo{collaboration}{LIGO
  Scientific, Virgo}),
\newblock \bibinfo{title}{{GW170817: Measurements of neutron star radii and
  equation of state}},
\newblock \bibinfo{journal}{Phys. Rev. Lett.} \bibinfo{volume}{121}
  (\bibinfo{year}{2018}) \bibinfo{pages}{161101}.
  \DOIprefix\doi{10.1103/PhysRevLett.121.161101}.
  \href{http://arxiv.org/abs/1805.11581}{{\tt arXiv:1805.11581}}.
\bibitem[{Newman and Penrose(1963)}]{newman_penrose_gw}
\bibinfo{author}{E.~Newman}, \bibinfo{author}{R.~Penrose},
\newblock \bibinfo{title}{Errata: An approach to gravitational radiation by a
  method of spin coefficients},
\newblock \bibinfo{journal}{Journal of Mathematical Physics}
  \bibinfo{volume}{4} (\bibinfo{year}{1963}) \bibinfo{pages}{998--998}.
  \URLprefix \url{https://doi.org/10.1063/1.1704025}.
  \DOIprefix\doi{10.1063/1.1704025}.
  \href{http://arxiv.org/abs/https://doi.org/10.1063/1.1704025}{{\tt
  arXiv:https://doi.org/10.1063/1.1704025}}.
\bibitem[{Bishop and Rezzolla(2016)}]{bishop_rezzolla_gw_extraction}
\bibinfo{author}{N.~T. Bishop}, \bibinfo{author}{L.~Rezzolla},
\newblock \bibinfo{title}{{Extraction of Gravitational Waves in Numerical
  Relativity}},
\newblock \bibinfo{journal}{Living Rev. Rel.} \bibinfo{volume}{19}
  (\bibinfo{year}{2016}) \bibinfo{pages}{2}.
  \DOIprefix\doi{10.1007/s41114-016-0001-9}.
  \href{http://arxiv.org/abs/1606.02532}{{\tt arXiv:1606.02532}}.
\bibitem[{Kastaun et~al.(2021)Kastaun, Kalinani, and Ciolfi}]{Kastaun-2021}
\bibinfo{author}{W.~Kastaun}, \bibinfo{author}{J.~V. Kalinani},
  \bibinfo{author}{R.~Ciolfi},
\newblock \bibinfo{title}{Robust recovery of primitive variables in
  relativistic ideal magnetohydrodynamics},
\newblock \bibinfo{journal}{Phys. Rev. D} \bibinfo{volume}{103}
  (\bibinfo{year}{2021}) \bibinfo{pages}{023018}. \URLprefix
  \url{https://link.aps.org/doi/10.1103/PhysRevD.103.023018}.
  \DOIprefix\doi{10.1103/PhysRevD.103.023018}.
\bibitem[{{Finn}(1988)}]{Finn_1988}
\bibinfo{author}{L.~S. {Finn}},
\newblock \bibinfo{title}{{Relativistic stellar pulsations in the Cowling
  approximation}},
\newblock \bibinfo{journal}{Mon. Not. Roy. Astron. Soc.} \bibinfo{volume}{232}
  (\bibinfo{year}{1988}) \bibinfo{pages}{259--275}.
  \DOIprefix\doi{10.1093/mnras/232.2.259}.
\bibitem[{Hinderer(2008)}]{Hinderer_2008}
\bibinfo{author}{T.~Hinderer},
\newblock \bibinfo{title}{{Tidal Love numbers of neutron stars}},
\newblock \bibinfo{journal}{Astrophys. J.} \bibinfo{volume}{677}
  (\bibinfo{year}{2008}) \bibinfo{pages}{1216--1220}.
  \DOIprefix\doi{10.1086/533487}. \href{http://arxiv.org/abs/0711.2420}{{\tt
  arXiv:0711.2420}}.
\bibitem[{Press et~al.(2007)Press, Teukolsky, Vetterling, and
  Flannery}]{NumRec}
\bibinfo{author}{W.~H. Press}, \bibinfo{author}{S.~A. Teukolsky},
  \bibinfo{author}{W.~T. Vetterling}, \bibinfo{author}{B.~P. Flannery},
  \bibinfo{title}{Numerical Recipes 3rd Edition: The Art of Scientific
  Computing}, \bibinfo{publisher}{Cambridge University Press},
  \bibinfo{year}{2007}.
\bibitem[{Lau et~al.(2010)Lau, Leung, and Lin}]{Lau:2010}
\bibinfo{author}{H.~K. Lau}, \bibinfo{author}{P.~T. Leung},
  \bibinfo{author}{L.~M. Lin},
\newblock \bibinfo{title}{{Inferring physical parameters of compact stars from
  their f-mode gravitational wave signals}},
\newblock \bibinfo{journal}{Astrophys. J.} \bibinfo{volume}{714}
  (\bibinfo{year}{2010}) \bibinfo{pages}{1234--1238}.
  \DOIprefix\doi{10.1088/0004-637X/714/2/1234}.
  \href{http://arxiv.org/abs/0911.0131}{{\tt arXiv:0911.0131}}.
\bibitem[{Baiotti et~al.(2009)Baiotti, Bernuzzi, Corvino, De~Pietri, and
  Nagar}]{baiotti_pert}
\bibinfo{author}{L.~Baiotti}, \bibinfo{author}{S.~Bernuzzi},
  \bibinfo{author}{G.~Corvino}, \bibinfo{author}{R.~De~Pietri},
  \bibinfo{author}{A.~Nagar},
\newblock \bibinfo{title}{{Gravitational-Wave Extraction from Neutron Stars
  Oscillations: Comparing linear and nonlinear techniques}},
\newblock \bibinfo{journal}{Phys. Rev. D} \bibinfo{volume}{79}
  (\bibinfo{year}{2009}) \bibinfo{pages}{024002}.
  \DOIprefix\doi{10.1103/PhysRevD.79.024002}.
  \href{http://arxiv.org/abs/0808.4002}{{\tt arXiv:0808.4002}}.
\bibitem[{{Deppe} et~al.(2021){Deppe}, {H{\'e}bert}, {Kidder}, {Throwe},
  {Anantpurkar}, {Armaza}, {Bonilla}, {Boyle}, {Chaudhary}, {Duez}, {Fischer},
  {Foucart}, {Giesler}, {Guo}, {Kim}, {Kumar}, {Legred}, {Li}, {Lovelace},
  {Ma}, {Macedo}, {Melchor}, {Morales}, {Moxon}, {Nelli}, {O'Shea}, {Pfeiffer},
  {Ramirez}, {R{\"u}ter}, {Sanchez}, {Scheel}, {Thomas}, {Vieira}, {Wittek},
  {Wlodarczyk}, and {Teukolsky}}]{Deppe-2021}
\bibinfo{author}{N.~{Deppe}}, \bibinfo{author}{F.~{H{\'e}bert}},
  \bibinfo{author}{L.~E. {Kidder}}, \bibinfo{author}{W.~{Throwe}},
  \bibinfo{author}{I.~{Anantpurkar}}, \bibinfo{author}{C.~{Armaza}},
  \bibinfo{author}{G.~S. {Bonilla}}, \bibinfo{author}{M.~{Boyle}},
  \bibinfo{author}{H.~{Chaudhary}}, \bibinfo{author}{M.~D. {Duez}},
  \bibinfo{author}{N.~L. {Fischer}}, \bibinfo{author}{F.~{Foucart}},
  \bibinfo{author}{M.~{Giesler}}, \bibinfo{author}{J.~S. {Guo}},
  \bibinfo{author}{Y.~{Kim}}, \bibinfo{author}{P.~{Kumar}},
  \bibinfo{author}{I.~{Legred}}, \bibinfo{author}{D.~{Li}},
  \bibinfo{author}{G.~{Lovelace}}, \bibinfo{author}{S.~{Ma}},
  \bibinfo{author}{A.~{Macedo}}, \bibinfo{author}{D.~{Melchor}},
  \bibinfo{author}{M.~{Morales}}, \bibinfo{author}{J.~{Moxon}},
  \bibinfo{author}{K.~C. {Nelli}}, \bibinfo{author}{E.~{O'Shea}},
  \bibinfo{author}{H.~P. {Pfeiffer}}, \bibinfo{author}{T.~{Ramirez}},
  \bibinfo{author}{H.~R. {R{\"u}ter}}, \bibinfo{author}{J.~{Sanchez}},
  \bibinfo{author}{M.~A. {Scheel}}, \bibinfo{author}{S.~{Thomas}},
  \bibinfo{author}{D.~{Vieira}}, \bibinfo{author}{N.~A. {Wittek}},
  \bibinfo{author}{T.~{Wlodarczyk}}, \bibinfo{author}{S.~A. {Teukolsky}},
\newblock \bibinfo{title}{{Simulating magnetized neutron stars with
  discontinuous Galerkin methods}},
\newblock \bibinfo{journal}{arXiv e-prints}  (\bibinfo{year}{2021})
  \bibinfo{pages}{arXiv:2109.12033}.
  \href{http://arxiv.org/abs/2109.12033}{{\tt arXiv:2109.12033}}.
\bibitem[{Pradhan and
  Chatterjee(2021)}]{pradhan_chatterjee2021:PhysRevC.103.035810}
\bibinfo{author}{B.~K. Pradhan}, \bibinfo{author}{D.~Chatterjee},
\newblock \bibinfo{title}{{Effect of hyperons on f-mode oscillations in Neutron
  Stars}},
\newblock \bibinfo{journal}{Phys. Rev. C} \bibinfo{volume}{103}
  (\bibinfo{year}{2021}) \bibinfo{pages}{035810}.
  \DOIprefix\doi{10.1103/PhysRevC.103.035810}.
  \href{http://arxiv.org/abs/2011.02204}{{\tt arXiv:2011.02204}}.
\bibitem[{{Wen} et~al.(2019){Wen}, {Li}, {Chen}, and {Zhang}}]{Wen-2019}
\bibinfo{author}{D.-H. {Wen}}, \bibinfo{author}{B.-A. {Li}},
  \bibinfo{author}{H.-Y. {Chen}}, \bibinfo{author}{N.-B. {Zhang}},
\newblock \bibinfo{title}{{GW170817 implications on the frequency and damping
  time of f -mode oscillations of neutron stars}},
\newblock \bibinfo{journal}{Phys. Rev. C} \bibinfo{volume}{99}
  (\bibinfo{year}{2019}) \bibinfo{pages}{045806}.
  \DOIprefix\doi{10.1103/PhysRevC.99.045806}.
  \href{http://arxiv.org/abs/1901.03779}{{\tt arXiv:1901.03779}}.
\bibitem[{Maurya et~al.(2022{\natexlab{a}})Maurya, Mustafa, Govender, and
  Newton~Singh}]{ref1}
\bibinfo{author}{S.~K. Maurya}, \bibinfo{author}{G.~Mustafa},
  \bibinfo{author}{M.~Govender}, \bibinfo{author}{K.~Newton~Singh},
\newblock \bibinfo{title}{{Exploring physical properties of minimally deformed
  strange star model and constraints on maximum mass limit in $f(\mathscr{Q})$
  gravity}},
\newblock \bibinfo{journal}{JCAP} \bibinfo{volume}{10}
  (\bibinfo{year}{2022}{\natexlab{a}}) \bibinfo{pages}{003}.
  \DOIprefix\doi{10.1088/1475-7516/2022/10/003}.
  \href{http://arxiv.org/abs/2207.02021}{{\tt arXiv:2207.02021}}.
\bibitem[{Maurya et~al.(2022{\natexlab{b}})Maurya, Singh, Govender, and
  Hansraj}]{ref2}
\bibinfo{author}{S.~K. Maurya}, \bibinfo{author}{K.~N. Singh},
  \bibinfo{author}{M.~Govender}, \bibinfo{author}{S.~Hansraj},
\newblock \bibinfo{title}{{Gravitationally Decoupled Strange Star Model beyond
  the Standard Maximum Mass Limit in
  Einstein\textendash{}Gauss\textendash{}Bonnet Gravity}},
\newblock \bibinfo{journal}{Astrophys. J.} \bibinfo{volume}{925}
  (\bibinfo{year}{2022}{\natexlab{b}}) \bibinfo{pages}{208}.
  \DOIprefix\doi{10.3847/1538-4357/ac4255}.
  \href{http://arxiv.org/abs/2109.00358}{{\tt arXiv:2109.00358}}.
\bibitem[{Deb et~al.(2019)Deb, Ketov, Maurya, Khlopov, Moraes, and Ray}]{ref3}
\bibinfo{author}{D.~Deb}, \bibinfo{author}{S.~V. Ketov}, \bibinfo{author}{S.~K.
  Maurya}, \bibinfo{author}{M.~Khlopov}, \bibinfo{author}{P.~H. R.~S. Moraes},
  \bibinfo{author}{S.~Ray},
\newblock \bibinfo{title}{{Exploring physical features of anisotropic strange
  stars beyond standard maximum mass limit in $f(R,\mathcal {T})$ gravity}},
\newblock \bibinfo{journal}{Mon. Not. Roy. Astron. Soc.} \bibinfo{volume}{485}
  (\bibinfo{year}{2019}) \bibinfo{pages}{5652--5665}.
  \DOIprefix\doi{10.1093/mnras/stz708}.
  \href{http://arxiv.org/abs/1810.07678}{{\tt arXiv:1810.07678}}.
\bibitem[{Maurya et~al.(2022)Maurya, Singh, Govender, and Ray}]{ref4}
\bibinfo{author}{S.~K. Maurya}, \bibinfo{author}{K.~N. Singh},
  \bibinfo{author}{M.~Govender}, \bibinfo{author}{S.~Ray},
\newblock \bibinfo{title}{{Observational constraints on maximum mass limit and
  physical properties of anisotropic strange star models by gravitational
  decoupling in Einstein\textendash{}Gauss\textendash{}Bonnet gravity}},
\newblock \bibinfo{journal}{Mon. Not. Roy. Astron. Soc.} \bibinfo{volume}{519}
  (\bibinfo{year}{2022}) \bibinfo{pages}{4303--4324}.
  \DOIprefix\doi{10.1093/mnras/stac3611}.
\bibitem[{Baker et~al.(2015)Baker, Psaltis, and Skordis}]{Baker:2015}
\bibinfo{author}{T.~Baker}, \bibinfo{author}{D.~Psaltis},
  \bibinfo{author}{C.~Skordis},
\newblock \bibinfo{title}{{Linking Tests of Gravity On All Scales: from the
  Strong-Field Regime to Cosmology}},
\newblock \bibinfo{journal}{Astrophys. J.} \bibinfo{volume}{802}
  (\bibinfo{year}{2015}) \bibinfo{pages}{63}.
  \DOIprefix\doi{10.1088/0004-637X/802/1/63}.
  \href{http://arxiv.org/abs/1412.3455}{{\tt arXiv:1412.3455}}.
\bibitem[{Godzieba et~al.(2021)Godzieba, Gamba, Radice, and
  Bernuzzi}]{Godzieba-etal:arxiv2012}
\bibinfo{author}{D.~A. Godzieba}, \bibinfo{author}{R.~Gamba},
  \bibinfo{author}{D.~Radice}, \bibinfo{author}{S.~Bernuzzi},
\newblock \bibinfo{title}{{Updated universal relations for tidal
  deformabilities of neutron stars from phenomenological equations of state}},
\newblock \bibinfo{journal}{Phys. Rev. D} \bibinfo{volume}{103}
  (\bibinfo{year}{2021}) \bibinfo{pages}{063036}.
  \DOIprefix\doi{10.1103/PhysRevD.103.063036}.
  \href{http://arxiv.org/abs/2012.12151}{{\tt arXiv:2012.12151}}.

\end{thebibliography}

\end{document}